\PassOptionsToPackage{dvipsnames}{xcolor}
\documentclass[sigconf,screen,nonacm]{acmart}

\usepackage{fontaxes}
\usepackage{multirow}
\usepackage{xspace}
\usepackage{xurl}
\usepackage{tikz}
\usetikzlibrary{arrows.meta}
\usepackage{pgfplots}
\pgfplotsset{compat=1.18}
\usepgfplotslibrary{groupplots}
\setcopyright{none}

\definecolor{PaperProtected}{HTML}{4E79A7}
\definecolor{PaperCipher}{HTML}{F28E2B}
\definecolor{PaperHighlight}{HTML}{E15759}
\definecolor{PaperInk}{HTML}{333333}
\definecolor{PaperBackup}{HTML}{59A14F}
\definecolor{PaperToken}{HTML}{7B68B5}

\ifdefined\pdfmapfile\pdfmapfile{+libertine.map}\fi
\newcommand{\figurefont}{\fontencoding{T1}\fontfamily{LinuxBiolinumT-TLF}}

\newcommand{\sysname}{\texorpdfstring{\textulc{\textit{HiPerfCC-\allowbreak StateReuse}}}{HiPerfCC-StateReuse}\xspace}
\newcommand{\laegis}{L\AE{}GIS\xspace}
\newcommand{\bheading}[1]{\noindent\textbf{#1}}
\newcommand{\stepnum}[1]{\ding{\numexpr181+#1\relax}}
\newcommand{\secref}[1]{\mbox{Section~\ref{#1}}\xspace}
\newcommand{\figref}[1]{\mbox{Figure~\ref{#1}}}
\newcommand{\tabref}[1]{\mbox{Table~\ref{#1}}}

\newenvironment{packeditemize}{
\begin{list}{$\bullet$}{
\setlength{\labelwidth}{0pt}
\setlength{\itemsep}{2pt}
\setlength{\leftmargin}{\labelwidth}
\addtolength{\leftmargin}{\labelsep}
\setlength{\parindent}{0pt}
\setlength{\listparindent}{\parindent}
\setlength{\parsep}{1pt}
\setlength{\topsep}{1pt}}}{\end{list}}

\title[\sysname: Ciphertext Reuse]{\sysname: Ciphertext Reuse\texorpdfstring{\\}{ }for Fast Confidential GPU Data Movement}

\author{Haoxuan Xu}
\email{xuhaoxua@usc.edu}
\affiliation{
  \institution{University of Southern California}
  \city{Los Angeles}
  \country{USA}
}

\author{Shaoyu Wang}
\email{wangshao@usc.edu}
\affiliation{
  \institution{University of Southern California}
  \city{Los Angeles}
  \country{USA}
}

\author{Minghao Lin}
\email{mlin7282@usc.edu}
\affiliation{
  \institution{University of Southern California}
  \city{Los Angeles}
  \country{USA}
}

\author{Beijie Liu}
\email{beijieli@usc.edu}
\affiliation{
  \institution{University of Southern California}
  \city{Los Angeles}
  \country{USA}
}

\author{Potung Yu}
\email{potungyu@usc.edu}
\affiliation{
  \institution{University of Southern California}
  \city{Los Angeles}
  \country{USA}
}

\author{Yongqin Wang}
\email{yongqin@usc.edu}
\affiliation{
  \institution{University of Southern California}
  \city{Los Angeles}
  \country{USA}
}

\author{Vincent Quentin Ulitzsch}
\email{viniul@mit.edu}
\affiliation{
  \institution{Massachusetts Institute of Technology}
  \city{Cambridge}
  \country{USA}
}

\author{Seo Jin Park}
\email{seojinpa@usc.edu}
\affiliation{
  \institution{University of Southern California}
  \city{Los Angeles}
  \country{USA}
}

\author{Murali Annavaram}
\email{annavara@usc.edu}
\affiliation{
  \institution{University of Southern California}
  \city{Los Angeles}
  \country{USA}
}

\author{Mengyuan Li}
\email{mli49061@usc.edu}
\affiliation{
  \institution{University of Southern California}
  \city{Los Angeles}
  \country{USA}
}

\renewcommand{\shortauthors}{Xu et al.}

\makeatletter
\renewcommand{\@mkauthors}{%
  \global\setbox\mktitle@bx=\vbox{%
  \unvbox\mktitle@bx\par\medskip
  \centering
  {\large\sffamily
    Haoxuan Xu\textsuperscript{1}\quad Shaoyu Wang\textsuperscript{1}\quad Minghao Lin\textsuperscript{1}\quad Beijie Liu\textsuperscript{1}\par
    Potung Yu\textsuperscript{1}\quad Yongqin Wang\textsuperscript{1}\quad Vincent Quentin Ulitzsch\textsuperscript{2}\par
    Seo Jin Park\textsuperscript{1}\quad Murali Annavaram\textsuperscript{1}\quad Mengyuan Li\textsuperscript{1}\par
  }
  \vspace{5pt}
  {\normalsize
    \textsuperscript{1}University of Southern California\quad
    \textsuperscript{2}Massachusetts Institute of Technology\par
  }
  \vspace{4pt}
  {\small\ttfamily
    \href{mailto:xuhaoxua@usc.edu}{xuhaoxua@usc.edu}\quad \href{mailto:wangshao@usc.edu}{wangshao@usc.edu}\quad \href{mailto:mlin7282@usc.edu}{mlin7282@usc.edu}\quad \href{mailto:beijieli@usc.edu}{beijieli@usc.edu}\par
    \href{mailto:potungyu@usc.edu}{potungyu@usc.edu}\quad \href{mailto:yongqin@usc.edu}{yongqin@usc.edu}\quad \href{mailto:viniul@mit.edu}{viniul@mit.edu}\par
    \href{mailto:seojinpa@usc.edu}{seojinpa@usc.edu}\quad \href{mailto:annavara@usc.edu}{annavara@usc.edu}\quad \href{mailto:mli49061@usc.edu}{mli49061@usc.edu}\par
  }
  \bigskip
  }%
}
\makeatother

\begin{document}

\begin{abstract}
Confidential GPUs offer protected GPU computation with near-native on-device compute performance, but data movement remains costly. Transferring application data or model weights between CPU and GPU TEEs through untrusted host memory requires bulk CPU encryption and decryption, even when the contents remain unchanged. Existing software techniques overlap or parallelize these operations but retain their per-byte CPU cost. Reusing exported ciphertext could avoid this work. However, GPU TEE firmware enforces direction-specific encryption keys and fresh per-transfer initialization vectors (IVs), preventing direct ciphertext reuse.

In this paper, we introduce ciphertext rebinding, a purely software primitive that enables repeated and cross-GPU ciphertext reuse on existing GPU TEEs without hardware changes, while preserving confidentiality, integrity, and freshness.
Specifically, ciphertext rebinding exploits the separation between ciphertext-dependent and IV-dependent components in GPU TEE's AES-GCM authentication: a GPU-precomputed summary under the destination key can enable the CPU to update authentication tags without processing the payload. With this modification, GPU's subsequent read on the retained ciphertext can still proceed through the GPU's hardware-fixed copy-engine interfaces, which cannot be modified by software. An additional GPU correction step can then compensate for differences in the keys and IVs to recover the original plaintext. By retaining GPU-exported ciphertext in untrusted host memory, ciphertext rebinding avoids bulk CPU decryption and re-encryption, improving effective bandwidth for GPU-state restoration and sharing.
We implement this primitive in \sysname through only driver and library changes, combining batched submissions with pipelined GPU processing and copy-engine transfers. Our evaluation measures transfer efficiency and CPU overhead through microbenchmarks and examines application performance through three representative case studies: model switching, key–value cache offloading, and cross-instance cache sharing.

\end{abstract}

\maketitle
\hypersetup{pdfauthor={Haoxuan Xu, Shaoyu Wang, Minghao Lin, Beijie Liu, Potung Yu, Yongqin Wang, Vincent Quentin Ulitzsch, Seo Jin Park, Murali Annavaram, Mengyuan Li}}

\section{Introduction}
\label{sec:introduction}
Confidential computing (CC) uses hardware-enforced trusted execution environments (TEEs) to protect sensitive data and computation from privileged infrastructure software. NVIDIA confidential GPUs extend this protection to GPU execution, with near-native performance for computation within protected GPU memory~\cite{ccoverview}. Moving data between a CPU confidential virtual machine (CVM) and a confidential GPU, however, remains expensive~\cite{dissect_cc,ispass_cc}. On existing GPU CC hardware, transfers that pass through untrusted shared host buffers are protected by authenticated encryption. The CVM encrypts host-to-GPU data and decrypts GPU-to-host data in software, while the confidential GPU copy engines (CEs) perform the corresponding decryption/encryption operations in hardware~\cite{ccoverview}.

This CPU cryptographic work is repeated even when GPU applications restore the unchanged GPU state from host memory. For example, model switching may repeatedly load the same weights, while hierarchical key--value (KV) caching offloads and reuses cached state for shared prefixes~\cite{vllmsleep,cachedattention,hicache}. Under confidential GPU's conventional protected transfers, backing up such state requires the CPU to decrypt GPU-produced ciphertext into private CVM memory. Every subsequent restoration requires the CPU to re-encrypt the same payload again. Although the state has not changed, the CPU must process redundant decryption and re-encryption of every byte on each restoration.

Prior approaches reduce protected-transfer overhead either by changing the hardware or protocol in simulation, or by making the existing CPU cryptographic work more efficient. \textit{Hardware-oriented designs} include \laegis{}, which adds an HBM-resident initialization-vector (IV) bank to decouple CPU pre-encryption from GPU page-migration order~\cite{laegis}. TensorTEE extends the CPU and NPU architectures so that their TEEs can exchange tensors without re-encryption~\cite{2024:tensortee:han}. Fastrack changes the GPU CC protocol to establish a direct client--GPU secure channel~\cite{fastrack}. 
These hardware-oriented designs require architectural or protocol modifications and thus cannot be directly deployed on existing GPU CC hardware. Instead, they represent potential directions for future hardware and protocol designs.
In contrast, \textit{software optimizations} target existing GPU CC hardware without requiring architectural or protocol changes. They aim to reduce protected-transfer overhead by hiding CPU cryptographic latency or parallelizing cryptographic operations. PipeLLM predicts which blocks will be restored and in what order. The CPU encrypts each block under the anticipated IV while the GPU computes~\cite{pipellm}. Even with perfect predictions, the CPU encrypts the full payload anew for every restoration. If the channel advances past the predicted IV, PipeLLM must discard the speculative ciphertext and can fall back to on-demand encryption. Serialized Bridge parallelizes protected transfers across CUDA contexts, retaining the per-byte CPU cryptographic work~\cite{serializedbridge}. FLASH retains GPU-produced ciphertext when creating checkpoints, but its recovery path decrypts the saved state inside the CVM and resumes conventional protected transfers~\cite{flash}. Consequently, these software techniques do not eliminate bulk CPU cryptography when unchanged GPU state is repeatedly restored.

\bheading{Enabling Efficient GPU State Reuse.} Reusing GPU-produced ciphertext would eliminate repeated CPU cryptographic work, but existing confidential GPU transfer protocols prevent direct reuse. Confidential GPU transfers use direction-specific keys and fresh IVs, so ciphertext from a GPU-to-host transfer cannot simply be replayed on the host-to-GPU channel. Its original AES-GCM authentication tag is invalid under the destination key and IV. Replacing the tag alone is also insufficient: even if the CE authenticates the retained ciphertext, decryption under the new key and IV yields a different plaintext. Reuse therefore requires solving two distinct problems: authenticating retained ciphertext for the destination transfer and recovering the original plaintext, both without bulk CPU cryptography. In this paper, we ask whether both problems can instead be solved in software through existing confidential GPU hardware.

We introduce \textbf{ciphertext rebinding}, a new, purely software primitive that enables repeated and cross-GPU state reuse on existing GPU TEEs without hardware changes. 
Ciphertext rebinding exploits the separation between ciphertext-dependent and IV-dependent components of AES-GCM authentication~\cite{gcm}. Specifically, a GPU precomputes a ciphertext-dependent authentication summary under the destination channel's key from trusted GPU data. For each subsequent host-to-GPU transfer, the CPU combines this retained summary with a fresh IV-dependent mask to update the authentication tag without processing the payload. The new tag lets the existing CE authenticate and decrypt the retained ciphertext under its current channel state. A GPU kernel then exploits the XOR structure of counter-mode encryption to compensate for the difference between the original GPU-to-host and current host-to-GPU keystreams, recovering the original plaintext. By retaining the ciphertext in untrusted shared host memory, ciphertext rebinding avoids bulk CPU decryption and re-encryption, improving effective bandwidth for GPU-state restoration and sharing. 
Ciphertext rebinding also preserves confidentiality, integrity, and freshness of the GPU state under GPU CC's threat model. We substantiate these guarantees through a security analysis based on standard AES-GCM bounds and trusted IV progression, complemented by machine-checked proofs of correction correctness and confidentiality and integrity reductions in EasyCrypt~\cite{easycrypt}.

We implement ciphertext rebinding in \sysname through only driver and library changes. GPU summary computation and plaintext correction still process the payload, and transfers incur submission and synchronization costs. \sysname batches submissions to amortize these costs and pipelines GPU processing with CE transfers to overlap the remaining work.
We evaluate \sysname through transfer microbenchmarks and three application case studies: model switching, KV-cache offloading, and cross-instance KV-cache sharing. Compared with Multi-context CC, a baseline that parallelizes conventional protected transfers across CUDA contexts~\cite{serializedbridge}, \sysname reduces CPU time per GiB by 92--97\% in microbenchmarks on H200 and RTX PRO 6000. On H200, it achieves 1.9--2.3$\times$ the baseline's effective transfer bandwidth, reaching 54--57~GB/s and approaching the bandwidth achieved with GPU CC disabled (CC-off).

This paper makes three contributions:
\begin{packeditemize}
\item We introduce ciphertext rebinding, a primitive that adapts authentication and plaintext recovery to the destination channel's key and fresh per-transfer IVs, enabling repeated and cross-GPU ciphertext reuse through existing CE interfaces without bulk CPU decryption or re-encryption. We analyze its confidentiality, integrity, and freshness under the stated threat model. We machine-check the correction identity and the confidentiality and integrity reductions in an abstract EasyCrypt model~\cite{easycrypt}.
\item We implement \sysname through driver and library changes that batch submissions and pipeline GPU processing with CE transfers on existing confidential GPUs, without GPU hardware or firmware changes. Together, batching and pipelining increase transfer bandwidth on both platforms by 23--60\% over using neither optimization.
\item We evaluate \sysname against Multi-context CC on H200. \sysname reduces complete model-restoration latency by 38--59\% across two models and two engines. It achieves 1.38$\times$ the baseline's output-token throughput with KV offloading (geometric mean over nine model--workload pairs at peak load) and 1.38$\times$ its throughput with cross-instance KV sharing.
\end{packeditemize}

\bheading{Scope of the system.} \sysname targets confidential GPU deployments that use software-encrypted host bounce buffers for CPU--GPU transfers. These include shipping Hopper and Blackwell passthrough modes and cloud offerings such as Google Cloud A3 H100, Azure H100, and VoltageGPU's RTX PRO 6000 confidential VMs~\cite{2026:trustedcomputingr595:nvidia,2024:confidentialaccelerators:young,2024:azurecc:fournet,2026:voltagegpu:aubry}. Thus, \sysname can reduce the CPU overhead of restoring and sharing GPU state across a broad range of existing confidential GPU deployments. Trusted I/O technologies can accelerate CPU--GPU communication by combining trusted device assignment through TDISP with hardware-accelerated inline encryption through PCIe IDE~\cite{2022:tdisp:pcisig,2022:ide:pcisig,2025:secureai:nvidia}. However, this approach requires compatible CPUs, GPUs, firmware, and drivers~\cite{2025:secureai:nvidia}, which are not yet publicly commercially available. TDISP integration therefore remains a direction for future work, particularly as PCIe~6.0 platforms become more widely deployed in the future~\cite{2025:gputravelling:zhao}.

\bheading{Code availability.} Our code is available at \url{https://github.com/sept-usc/HiPerfCC-StateReuse}.

\section{Ciphertext Rebinding}
\label{sec:retagging}

Ciphertext rebinding keeps GPU-produced ciphertext unchanged while adapting authentication and plaintext recovery to the destination channel's key and fresh per-transfer IVs. The primitive combines tag updates with GPU plaintext correction to reuse this ciphertext through the existing copy-engine interface.

\subsection{Threat model} 
We follow the conventional threat model for CPU--GPU confidential computing~\cite{gpucc}. The host OS, hypervisor, shared host memory, and CPU--GPU interconnect are untrusted. An adversary controlling this infrastructure can observe, modify, inject, or replay data in shared memory and traffic on the interconnect. The trusted computing base includes the CPU/GPU security hardware and firmware and all software inside the CVM, including the guest OS, GPU drivers, applications, and our \sysname library and driver extensions. GPU kernels launched by this trusted software are also trusted. We assume successful attestation of the CPU and GPU environments and secure establishment of their communication keys. Under these assumptions, \sysname aims to preserve the confidentiality, integrity, and freshness of retained GPU state during storage in shared host memory and transfers across the CPU--GPU interconnect.

Access-pattern leakage is outside our security scope, consistent with the documented limitations of existing CPU--GPU confidential computing platforms. NVIDIA explicitly defers CPU-side side-channel protections to the underlying CPU TEE's threat model~\cite{nvidiathreatmodel}, AMD SEV-SNP excludes page-level access-pattern monitoring~\cite{sevsnp}, and Intel acknowledges residual memory-access-pattern leakage in TDX~\cite{inteltdxsidechannel}. Accordingly, \sysname does not attempt to hide access patterns, including the timing of state transfers or repeated imports of the same encrypted state. Other side-channel attacks, such as timing and cache attacks~\cite{timeslicesandwich,behindbars}, and denial-of-service (DoS) attacks that exhaust resources or prevent the CVM or GPU from making progress~\cite{teerealtime,notrefreshing}, are also outside the scope of this work. These exclusions are consistent with the GPU CC threat model~\cite{gpucc,nvidiathreatmodel}.

\subsection{Constraints on Existing GPU TEE}
\label{sec:cc-data-movement}

Existing GPU TEE hardware enforces direction-specific encryption keys and a fresh IV per transfer, preventing direct ciphertext reuse.

\bheading{Protected transfers.} Transfers between CVM-private memory and protected GPU memory pass through untrusted shared host buffers~\cite{gpucc}. For a host-to-device (H2D) transfer, the driver encrypts plaintext from CVM-private memory into a shared buffer. The CE reads the ciphertext over PCIe, verifies its authentication tag, and decrypts it into protected GPU memory. For a device-to-host (D2H) transfer, the CE writes encrypted GPU data to a shared buffer, and the CPU verifies and decrypts this ciphertext into CVM-private memory. Backing up GPU state as plaintext in CVM-private memory and later restoring it therefore requires CPU decryption followed by re-encryption of the payload.

\bheading{Directional channel state.} AES-GCM requires IV uniqueness for encryption under a given key~\cite{gcm}. H2D and D2H use different encryption keys and maintain separate IV sequences. Within each direction, the driver and CE must agree on the key and IV used by each transfer, denoted $(K_{\mathrm{H2D}},\mathrm{IV}_{\mathrm{H2D}})$ for H2D and $(K_{\mathrm{D2H}},\mathrm{IV}_{\mathrm{D2H}})$ for D2H. The driver tracks each channel's IV sequence as transfers are submitted, including the D2H IV needed for CPU decryption after the CE completes. The CE authenticates and decrypts H2D data using the channel's key and current IV through a fixed hardware interface~\cite{gpucc}. Reusing D2H ciphertext on the H2D path therefore requires a valid authentication tag under the H2D key and fresh IV. It also requires recovering the original plaintext despite the CE decrypting the retained ciphertext under that different key and IV.

\subsection{Tag Updates and Plaintext Correction}
\label{sec:key-insight}

\begin{figure}[t]
\centering
\begin{tikzpicture}[
    font=\figurefont\fontsize{8}{9.6}\selectfont,
    line width=.6pt,
    >={Latex[length=1.5mm,width=1.1mm]},
    box/.style={draw=black, line width=.6pt, rounded corners=.8pt,
        minimum height=4.6mm, inner xsep=2.5pt, inner ysep=1.2pt,
        align=center, fill=white},
    protected/.style={box, fill=PaperProtected!16},
    cipher/.style={box, fill=PaperCipher!20},
    neutral/.style={box, fill=black!6},
    zone/.style={draw=black, line width=.6pt, rounded corners=1.2pt},
    zonename/.style={font=\figurefont\bfseries\fontsize{8}{9.6}\selectfont,
        inner sep=1.5pt},
    panel/.style={font=\figurefont\bfseries\fontsize{8}{9.6}\selectfont,
        inner sep=0pt, anchor=base west},
    note/.style={font=\figurefont\fontsize{7}{8.4}\selectfont, inner sep=1pt},
    flow/.style={->, draw=black, line width=.6pt},
    xor/.style={draw=black, circle, line width=.6pt, fill=white,
        minimum size=3.2mm, inner sep=0pt},
    op/.style={minimum width=13mm, text height=1.5ex, text depth=.25ex},
    oplus/.style={xor, path picture={\draw
        (path picture bounding box.north) -- (path picture bounding box.south)
        (path picture bounding box.west) -- (path picture bounding box.east);}},
    val/.style={inner sep=1pt, text height=1.5ex, text depth=.25ex},
    input/.style={inner sep=0pt, outer sep=0pt, anchor=base},
    lbl/.style={midway, above, inner sep=1.3pt,
        text height=1.6ex, text depth=.55ex},
]
    \node[protected, op, minimum width=9mm] (ctr) at (1.90,0) {CTR};
    \node[oplus] (x1) at (3.12,0) {};
    \node[cipher, op] (ghash) at (4.54,0) {GHASH};
    \node[oplus] (x2) at (5.96,0) {};
    \node[val, anchor=east] (kiv) at (1.10,0) {$K,\mathrm{IV}$};
    \node[val, anchor=west] (tag) at (6.48,0) {$T$};
    \node[input] (p) at (3.12,.58) {$P$};
    \node[input] (h) at (4.54,.58) {$H$};
    \draw[flow] (kiv) -- (ctr);
    \draw[flow] (ctr) -- node[lbl] {$\mathrm{KS}$} (x1);
    \draw[flow] ([yshift=-1pt]p.south) -- (x1);
    \draw[flow] (x1) -- node[lbl] {$C$} (ghash);
    \draw[flow] ([yshift=-1pt]h.south) -- (ghash);
    \draw[flow] (ghash) -- node[lbl] {$g$} (x2);
    \draw[flow] (x2) -- (tag);

    \node[protected, op] (mask) at (4.54,-.68) {Tag mask};
    \node[val, anchor=east] (kiv2) at (3.54,-.68) {$K,\mathrm{IV}$};
    \draw[flow] (kiv2) -- (mask);
    \draw[flow] (mask) -| (x2);
\end{tikzpicture}
\caption{AES-GCM data dependencies. AAD and length encoding are omitted.}
\label{fig:cc-gcm}
\end{figure}
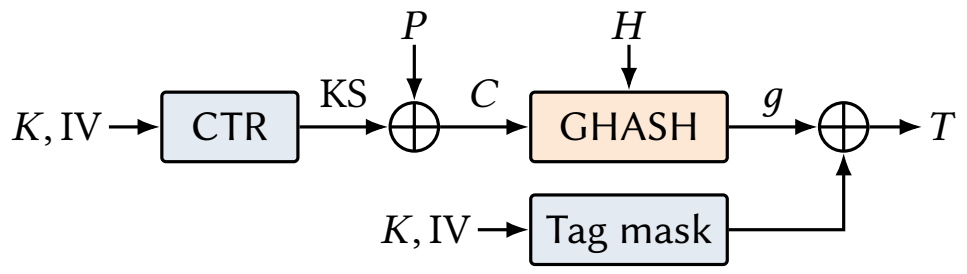

Ciphertext rebinding combines two steps: updating the authentication tag so the CE can accept the retained ciphertext, and correcting the CE output on the GPU to recover the original plaintext. Both steps exploit the structure of AES-GCM, illustrated in \figref{fig:cc-gcm}, while leaving the retained ciphertext and CE interface unchanged.

\bheading{AES-GCM structure.} Both endpoints use AES-GCM for authenticated encryption~\cite{gcm,gpucc}. We describe the primitive for one independently authenticated transfer segment. For a key $K$, initialization vector $\mathrm{IV}$, and plaintext $P$, counter-mode encryption produces ciphertext
\begin{equation}
    C = P \oplus \mathrm{KS}(K,\mathrm{IV}),
    \label{eq:gcm-ciphertext}
\end{equation}
where $\mathrm{KS}(K,\mathrm{IV})$ is the AES-generated keystream. For a full-length, 128-bit tag, authentication takes the form
\begin{equation}
    T = g_K(A,C) \oplus E_K(J_0(\mathrm{IV})).
    \label{eq:gcm-tag}
\end{equation}
Here $A$ is additional authenticated data (AAD), $E_K$ is AES encryption of one block, and $J_0(\mathrm{IV})$ is GCM's initial counter block derived from the IV. The term $g_K(A,C)$ is the GHASH result under $H=E_K(0^{128})$, including the prescribed padding and length encoding. The remaining term $E_K(J_0(\mathrm{IV}))$ is the IV-dependent tag mask. The bulk-data transfers studied in this paper use empty AAD ($A=\epsilon$). We use $\mathrm{KS}$ and $g$ as shorthand for $\mathrm{KS}(K,\mathrm{IV})$ and $g_K(A,C)$ when their inputs are clear from context.

\bheading{Rebinding authentication.} For fixed $K$, $A$, and $C$, the GHASH result in Equation~\eqref{eq:gcm-tag} is independent of the IV. Ciphertext rebinding exploits this separation by precomputing an \emph{authentication summary}, the ciphertext-dependent GHASH result under the destination H2D key:
\begin{equation}
    g_{\mathrm{H2D}} = g_{K_{\mathrm{H2D}}}(\epsilon,C).
    \label{eq:retag-summary}
\end{equation}
Although $C$ was produced under $K_{\mathrm{D2H}}$, its summary must use $K_{\mathrm{H2D}}$ to match the key used by the CE for H2D authentication. A GPU precomputes this summary from trusted data in protected GPU memory. This summary remains valid across H2D transfers as long as $C$ and $K_{\mathrm{H2D}}$ remain unchanged. On each subsequent H2D transfer, the CPU combines the retained summary with the tag mask for the channel's fresh $\mathrm{IV}_{\mathrm{H2D}}$:
\begin{equation}
    T_{\mathrm{new}} = g_{\mathrm{H2D}} \oplus
        E_{K_{\mathrm{H2D}}}(J_0(\mathrm{IV}_{\mathrm{H2D}})).
    \label{eq:retag-tag}
\end{equation}
Each tag update, or \emph{re-tagging}, thus requires only one AES-block encryption and one 128-bit XOR per segment, without CPU processing of the payload.

\bheading{Recovering plaintext after CE decryption.} The new tag allows the CE to authenticate the retained ciphertext, but decryption under the H2D key and IV produces an intermediate value rather than the original plaintext. Let $\mathrm{KS}_{\mathrm{D2H}}$ and $\mathrm{KS}_{\mathrm{H2D}}$ denote the keystreams for the original D2H and current H2D key/IV pairs, respectively. For stored ciphertext $C=P\oplus\mathrm{KS}_{\mathrm{D2H}}$, the CE writes the following intermediate value into protected GPU memory:
\begin{equation}
    P' = C \oplus \mathrm{KS}_{\mathrm{H2D}}
       = P \oplus \mathrm{KS}_{\mathrm{D2H}}
           \oplus \mathrm{KS}_{\mathrm{H2D}}.
    \label{eq:retag-intermediate}
\end{equation}
A GPU kernel exploits the XOR structure of counter mode to cancel both keystreams in the authenticated intermediate value:
\begin{equation}
    P = P' \oplus \mathrm{KS}_{\mathrm{H2D}}
           \oplus \mathrm{KS}_{\mathrm{D2H}}.
    \label{eq:retag-correction}
\end{equation}

\subsection{Ciphertext Export and Import}
\label{sec:export-import}

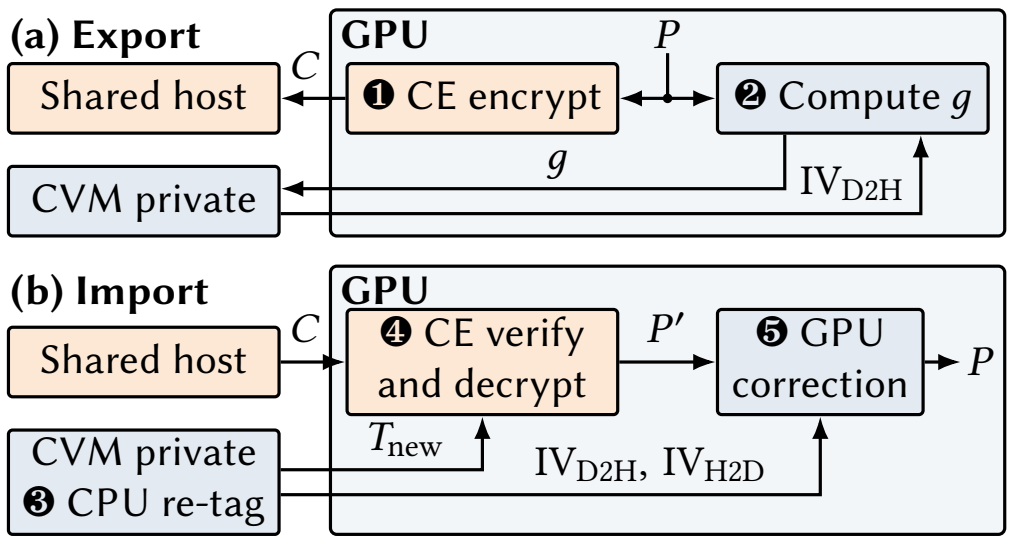
\begin{figure}[t]
\centering
\begin{tikzpicture}[
    font=\figurefont\fontsize{8}{9.6}\selectfont,
    line width=.6pt,
    >={Latex[length=1.5mm,width=1.1mm]},
    box/.style={draw=black, line width=.6pt, rounded corners=.8pt,
        minimum height=4.6mm, inner xsep=2.5pt, inner ysep=1.2pt,
        align=center, fill=white},
    protected/.style={box, fill=PaperProtected!16},
    cipher/.style={box, fill=PaperCipher!20},
    neutral/.style={box, fill=black!6},
    zone/.style={draw=black, line width=.6pt, rounded corners=1.2pt},
    zonename/.style={font=\figurefont\bfseries\fontsize{8}{9.6}\selectfont,
        inner sep=1.5pt},
    panel/.style={font=\figurefont\bfseries\fontsize{8}{9.6}\selectfont,
        inner sep=0pt, anchor=base west},
    note/.style={font=\figurefont\fontsize{7}{8.4}\selectfont, inner sep=1pt},
    flow/.style={->, draw=black, line width=.6pt},
    xor/.style={draw=black, circle, line width=.6pt, fill=white,
        minimum size=3.2mm, inner sep=0pt},
    col/.style={minimum width=17.9mm},
    one/.style={text height=1.6ex, text depth=.55ex},
    two/.style={minimum height=6.9mm, text depth=.55ex},
    lbl/.style={midway, above, inner sep=1.3pt,
        text height=1.6ex, text depth=.55ex},
    input/.style={inner sep=0pt, outer sep=0pt, anchor=base},
    val/.style={inner sep=1pt, text height=1.5ex, text depth=.25ex},
]
    \coordinate (zl) at (2.13,0);  %
    \draw[zone, fill=PaperProtected!7] (2.13,-.91) rectangle (6.59,.585);
    \node[zonename, anchor=north west] (gza) at (2.13,.585) {GPU};
    \node[panel] at (0,0 |- gza.base) {(a) Export};
    \node[cipher, col, one] (hostA) at (.895,0) {Shared host};
    \node[protected, col, one] (cvmA) at (.895,-.68) {CVM private};
    \node[cipher, col, one] (enc) at (3.135,0) {\stepnum{1} CE encrypt};
    \node[protected, col, one] (sum) at (5.585,0) {\stepnum{2} Compute $g$};
    \node[input] (pA) at (4.36,0 |- gza.base) {$P$};
    \draw ([yshift=-1pt]pA.south) -- (4.36,0);
    \draw[flow] (4.36,0) -- (enc);
    \draw[flow] (4.36,0) -- (sum);
    \fill (4.36,0) circle (.9pt);
    \draw[flow] (enc) -- (hostA);
    \path (hostA.east) -- node[lbl] {$C$} (zl);
    \coordinate (gA) at ([yshift=.825mm]cvmA.east);
    \coordinate (ivA) at ([yshift=-.825mm]cvmA.east);
    \coordinate (rgA) at ([xshift=-4.5mm]sum.south);
    \coordinate (riA) at ([xshift=4.5mm]sum.south);
    \draw[flow] (rgA) |- (gA);
    \path (gA -| zl) -- node[lbl] {$g$} (gA -| rgA);
    \draw[flow] (ivA) -| (riA);
    \path (ivA -| rgA) -- node[lbl] {$\mathrm{IV}_{\mathrm{D2H}}$} (ivA -| riA);

    \begin{scope}[yshift=-1.744cm]
    \draw[zone, fill=PaperProtected!7] (2.13,-1.14) rectangle (6.59,.634);
    \node[zonename, anchor=north west] (gzb) at (2.13,.634) {GPU};
    \node[panel] at (0,0 |- gzb.base) {(b) Import};
    \node[cipher, col, one] (hostB) at (.895,0) {Shared host};
    \node[protected, col, two] (cvmB) at (.895,-.795)
        {CVM private\\\stepnum{3} CPU re-tag};
    \node[cipher, col, two] (dec) at (3.135,0) {\stepnum{4} CE verify\\and decrypt};
    \node[protected, two, minimum width=13.6mm] (fix) at (5.37,0)
        {\stepnum{5} GPU\\correction};
    \node[val, anchor=west] (pB) at (6.30,0) {$P$};
    \draw[flow] (hostB) -- (dec);
    \path (hostB.east) -- node[lbl] {$C$} (zl |- hostB.east);
    \draw[flow] (dec) -- node[lbl] {$P'$} (fix);
    \draw[flow] (fix) -- (pB);
    \coordinate (tB) at ([yshift=.825mm]cvmB.east);
    \coordinate (ivB) at ([yshift=-.825mm]cvmB.east);
    \draw[flow] (tB) -| (dec.south);
    \path (tB -| zl) -- node[lbl] {$T_{\mathrm{new}}$} (tB -| dec.south);
    \draw[flow] (ivB) -| (fix.south);
    \path (ivB -| dec.south)
        -- node[lbl] {$\mathrm{IV}_{\mathrm{D2H}},\,\mathrm{IV}_{\mathrm{H2D}}$}
        (ivB -| fix.south);
    \end{scope}
\end{tikzpicture}
\caption{Ciphertext export and import. Here $g$ denotes
$g_{\mathrm{H2D}}$ and returns through the protected D2H path.
Key provisioning and completion checks are omitted.}
\label{fig:export-import}
\end{figure}

\emph{Export} stores GPU-produced ciphertext in shared host memory and prepares trusted metadata for subsequent reuse. Each \emph{import} uses this metadata to update the authentication tag and recover the original plaintext through CE decryption and GPU correction. \figref{fig:export-import} shows the operations for one authenticated segment on the same GPU.

\bheading{Trusted state.} The bulk ciphertext $C$ resides in untrusted shared host memory. For each exported segment, the library's CVM-private segment table retains $g_{\mathrm{H2D}}$, the original $\mathrm{IV}_{\mathrm{D2H}}$, and references to the source D2H and destination H2D keys. The table binds this state to the corresponding exported segment. The established keys remain in trusted channel state inside the CVM and GPU.

\bheading{Ciphertext export.} The application data to be exported may reside in noncontiguous GPU memory regions. The library first gathers these regions into a protected GPU staging buffer organized into transfer segments. When submitting a D2H transfer, the driver records its assigned $\mathrm{IV}_{\mathrm{D2H}}$. The CE encrypts the staged plaintext and writes $C$ directly to the shared host buffer (\stepnum{1} in \figref{fig:export-import}(a)). The library sends the recorded IV to the GPU through the protected H2D path. Using the same staged plaintext, D2H key, and recorded IV as the CE, a GPU kernel reproduces $C$ internally while computing its summary under $K_{\mathrm{H2D}}$ (\stepnum{2}), as defined in Equation~\eqref{eq:retag-summary}. The staged plaintext must remain unchanged until both the CE write and summary computation complete, ensuring that the summary corresponds to the ciphertext written by the CE.

The 128-bit summary returns to the segment's CVM-private metadata through the conventional protected D2H path. The library makes the exported segment available for import only after both the CE write and the protected summary readback complete.

\bheading{Ciphertext import.} To import a segment, the CPU retrieves its summary and the D2H key reference and IV recorded at export from the private segment table. The driver allocates a fresh $\mathrm{IV}_{\mathrm{H2D}}$ through the channel's normal progression. The CPU uses this IV and the retained $g_{\mathrm{H2D}}$ to compute $T_{\mathrm{new}}$ using Equation~\eqref{eq:retag-tag} (\stepnum{3} in \figref{fig:export-import}(b)). The CE reads the retained $C$, verifies the new tag, and decrypts into a protected GPU staging buffer, yielding $P'$ (\stepnum{4}). The library sends the recorded D2H IV and the newly allocated H2D IV to the GPU through the protected H2D path. A GPU kernel then applies Equation~\eqref{eq:retag-correction} to $P'$ using the export-time D2H key/IV and this transfer's H2D key/IV, recovering the original plaintext (\stepnum{5}). A scatter kernel then copies the recovered plaintext from staging to the application's destination regions. Import completes only after successful CE authentication, plaintext correction, and scatter.

Thus, the same ciphertext and trusted metadata support repeated imports, each using a fresh $\mathrm{IV}_{\mathrm{H2D}}$ and a newly computed tag.

\subsection{Ciphertext Rebinding Across GPUs}
\label{sec:retag-across-gpus}

Ciphertext rebinding also lets a destination GPU $B$ import ciphertext exported by a source GPU $A$ within the same CVM trust domain. The retained ciphertext $C$ remains encrypted under $A$'s D2H key and IV, while authentication must match $B$'s H2D channel.

Through protected CPU--GPU transfers, the trusted CVM supplies $A$ with $B$'s H2D GHASH key material to compute $g_{\mathrm{H2D},B}$ during export, and supplies $B$ with $A$'s D2H encryption key for plaintext correction. This key material is retained in trusted memory, while CVM-private metadata binds each segment's $g_{\mathrm{H2D},B}$ and original D2H IV to the corresponding source and destination channels. Import follows \secref{sec:export-import}, using $g_{\mathrm{H2D},B}$ and $B$'s fresh H2D IV for authentication, and $A$'s export-time D2H key/IV with $B$'s current H2D key/IV for correction.

For reuse by multiple destinations, export prepares a separate summary for each distinct destination H2D key, including $A$'s own H2D key when local re-import is needed. All destinations share the same stored $C$ and original D2H IV, while each uses its corresponding summary and a fresh IV from its own H2D channel.

\subsection{Batching and Pipelining}
\label{sec:pcie-speed}

The operations above avoid bulk CPU decryption and re-encryption, but still incur CE submission overhead and GPU work for gathering, summary computation, correction, and scattering. \sysname batches copy commands to amortize submission costs and pipelines CE transfers with this GPU work. We implement these operations through a C++/CUDA library and extensions to NVIDIA's Unified Virtual Memory (UVM) driver.

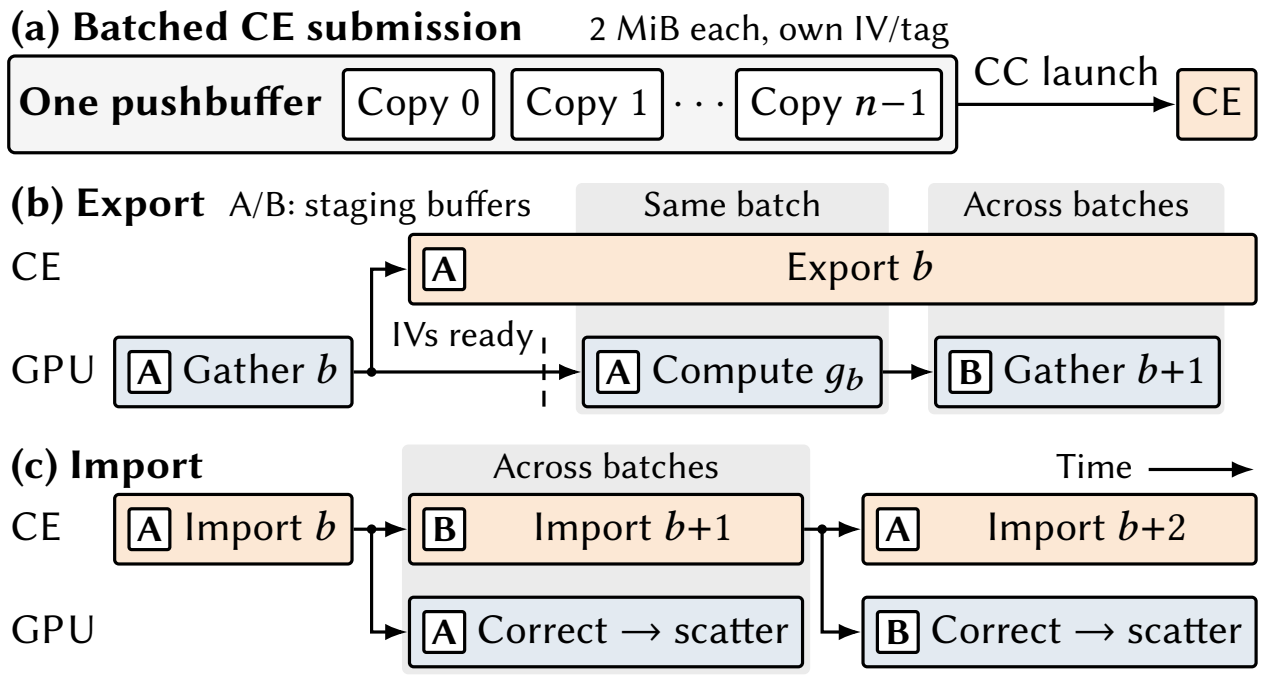
\begin{figure}[t]
\centering
\begin{tikzpicture}[
    font=\figurefont\fontsize{8}{9.6}\selectfont,
    line width=.6pt,
    >={Latex[length=1.5mm,width=1.1mm]},
    box/.style={draw=black, line width=.6pt, rounded corners=.8pt,
        minimum height=4.6mm, inner xsep=2.5pt, inner ysep=1.2pt,
        align=center, fill=white},
    protected/.style={box, fill=PaperProtected!16},
    cipher/.style={box, fill=PaperCipher!20},
    neutral/.style={box, fill=black!6},
    zone/.style={draw=black, line width=.6pt, rounded corners=1.2pt},
    zonename/.style={font=\figurefont\bfseries\fontsize{8}{9.6}\selectfont,
        inner sep=1.5pt},
    panel/.style={font=\figurefont\bfseries\fontsize{8}{9.6}\selectfont,
        inner sep=0pt, anchor=base west},
    note/.style={font=\figurefont\fontsize{7}{8.4}\selectfont, inner sep=1pt},
    flow/.style={->, draw=black, line width=.6pt},
    xor/.style={draw=black, circle, line width=.6pt, fill=white,
        minimum size=3.2mm, inner sep=0pt},
    one/.style={text height=1.6ex, text depth=.55ex},
    buf/.style={execute at begin node={\hspace*{3.57mm}}},
    key/.style={draw=black, fill=white, line width=.6pt, rounded corners=.5pt,
        minimum width=2.7mm, minimum height=2.7mm, inner sep=0pt,
        anchor=west, font=\figurefont\bfseries\fontsize{7}{8.4}\selectfont},
    lane/.style={one, anchor=west, inner sep=0pt},
    overlap/.style={fill=black!8, rounded corners=1.2pt},
    lbl/.style={midway, above, inner sep=1.3pt,
        text height=1.6ex, text depth=.55ex},
]
    \node[panel] at (0,.425) {(a) Batched CE submission};
    \node[note, anchor=base east] at (6.27,.425) {2 MiB each, own IV/tag};
    \draw[zone, fill=black!4] (0,-.32) rectangle (6.27,.32);
    \node[box, one] (c0) at (2.7105,0) {Copy $0$};
    \node[box, one] (c1) at (3.8265,0) {Copy $1$};
    \node[box, one] (cn) at (5.4915,0) {Copy $n{-}1$};
    \node[anchor=base, inner sep=0pt] at (4.591,0 |- c1.base) {$\cdots$};
    \node[zonename, anchor=base west] at (0,0 |- c0.base) {One pushbuffer};
    \node[cipher, one] (ce) at (7.991,0) {CE};
    \draw[flow] (6.27,0) -- node[lbl] {CC launch} (ce);

    \begin{scope}[yshift=-1.09cm]
    \fill[overlap] (3.751,-.96) rectangle (5.821,.57);
    \fill[overlap] (6.081,-.96) rectangle (8.041,.57);
    \node[panel] at (0,.35) {(b) Export};
    \node[note, anchor=base west] at (1.41,.35) {A/B: staging buffers};
    \node[note, anchor=base] at (4.786,.35) {Same batch};
    \node[note, anchor=base] at (7.061,.35) {Across batches};
    \node[lane] at (0,0) {CE};
    \node[lane] at (0,-.68) {GPU};

    \node[protected, one, buf, anchor=west, minimum width=15.7mm]
        (gather0) at (.69,-.68) {Gather $b$};
    \node[cipher, one, buf, anchor=west, minimum width=56mm]
        (export0) at (2.64,0) {Export $b$};
    \node[protected, one, buf, anchor=west, minimum width=19.7mm]
        (summary0) at (3.79,-.68) {Compute $g_b$};
    \node[protected, one, buf, anchor=west, minimum width=18.6mm]
        (gather1) at (6.12,-.68) {Gather $b{+}1$};
    \node[key] at ([xshift=.95mm]gather0.west) {A};
    \node[key] at ([xshift=.95mm]export0.west) {A};
    \node[key] at ([xshift=.95mm]summary0.west) {A};
    \node[key] at ([xshift=.95mm]gather1.west) {B};

    \coordinate (fb) at (2.40,-.68);
    \draw (gather0.east) -- (fb);
    \draw[flow] (fb) |- (export0.west);
    \draw[flow] (fb) -- (summary0.west);
    \fill (fb) circle (.9pt);
    \draw[dash pattern=on 3pt off 2pt] (3.54,-.68) ++(0,-2.293mm) -- ++(0,4.586mm);
    \node[note, anchor=base east] at (3.523,-.51) {IVs ready};
    \draw[flow] (summary0) -- (gather1);
    \end{scope}

    \begin{scope}[yshift=-2.81cm]
    \fill[overlap] (2.601,-.96) rectangle (5.301,.56);
    \node[panel] at (0,.34) {(c) Import};
    \node[note, anchor=base] at (3.951,.34) {Across batches};
    \node[note, anchor=base east] at (7.48,.34) {Time};
    \draw[flow] (7.54,.395) -- (8.24,.395);
    \node[lane] at (0,0) {CE};
    \node[lane] at (0,-.68) {GPU};

    \node[cipher, one, buf, anchor=west, minimum width=15.7mm]
        (import0) at (.69,0) {Import $b$};
    \node[cipher, one, buf, anchor=west, minimum width=26mm]
        (import1) at (2.64,0) {Import $b{+}1$};
    \node[cipher, one, buf, anchor=west, minimum width=26mm]
        (import2) at (5.64,0) {Import $b{+}2$};
    \node[protected, one, buf, anchor=west, minimum width=26mm]
        (process0) at (2.64,-.68) {Correct $\to$ scatter};
    \node[protected, one, buf, anchor=west, minimum width=26mm]
        (process1) at (5.64,-.68) {Correct $\to$ scatter};
    \node[key] at ([xshift=.95mm]import0.west) {A};
    \node[key] at ([xshift=.95mm]import1.west) {B};
    \node[key] at ([xshift=.95mm]import2.west) {A};
    \node[key] at ([xshift=.95mm]process0.west) {A};
    \node[key] at ([xshift=.95mm]process1.west) {B};

    \coordinate (f0) at (2.40,0);
    \coordinate (f1) at (5.38,0);
    \draw (import0.east) -- (f0);
    \draw[flow] (f0) -- (import1.west);
    \draw[flow] (f0) |- (process0.west);
    \fill (f0) circle (.9pt);
    \draw (import1.east) -- (f1);
    \draw[flow] (f1) -- (import2.west);
    \draw[flow] (f1) |- (process1.west);
    \fill (f1) circle (.9pt);
    \end{scope}
\end{tikzpicture}%
\caption{Batching and pipelining. Shading marks possible CE/kernel overlap.
The IV marker in (b) denotes delivery of the batch's recorded D2H IVs to
the GPU.}
\label{fig:batching-pipelining}
\end{figure}

\bheading{Batched CE submission.} Our implementation uses 2~MiB authenticated segments, matching UVM's maximum block size~\cite{nvidiaopen}. The UVM driver encodes CE copy commands in a command buffer, or \emph{pushbuffer}, for submission to a CE channel. Under CC, the driver encrypts this buffer, and the protected launch path authenticates and decrypts it before execution~\cite{gpucc,nvidiauvm}. Each submission therefore requires command preparation, protected launch, and completion tracking even when the bulk ciphertext is reused.

The library submits a batch of segments in one driver call, and the driver groups their copy commands into CE submissions, as shown in \figref{fig:batching-pipelining}(a). This amortizes driver-entry and protected-launch overhead across multiple segments, while each segment retains its own IV and authentication tag.

\bheading{Pipelined export.} After submitting a batch, the library sends its recorded D2H IVs to the GPU. Summary computation can then overlap the batch's remaining CE transfers (\figref{fig:batching-pipelining}(b)). Both operations read the same plaintext. Double buffering further overlaps gathering the next batch into buffer B with CE export from buffer A. A buffer is reused only after both its CE and GPU readers finish.

\bheading{Pipelined import.} Import uses separate staging storage. Within each batch, CE authentication and decryption, correction, and scatter execute in order. Each CE alternates between two buffers, overlapping import into B with correction and scatter of the preceding batch in A (\figref{fig:batching-pipelining}(c)). A is reusable only after scatter completes; import completes only after the driver confirms successful CE execution and the GPU finishes processing.

\sysname further exploits the GPU's multiple CEs by distributing independent exports across D2H channels and imports across H2D channels. Each import uses the authentication summary and H2D key/IV state corresponding to its selected channel.

\section{Security Analysis}
\label{sec:security}

In this section, we analyze the security of the ciphertext rebinding primitive. We examine plaintext confidentiality, integrity, and channel freshness during repeated imports of an exported ciphertext. 

\subsection{Confidentiality}
\label{sec:security-confidentiality}

\bheading{Property.} An adversary controlling shared host memory and the interconnect learns no plaintext information beyond segment lengths, the timing and order of transfers, and which stored ciphertext an import reuses. This guarantee covers both storage in shared memory and repeated imports.

\bheading{Argument.} Import tags and verification outcomes can be generated from the exported ciphertext and destination key, without the plaintext or source key. The adversary observes the exported ciphertext and its original D2H tag, the tags generated for imports, and whether verification succeeds. The ciphertext is produced by an ordinary D2H CC transfer under a fresh IV, so the source channel's encryption protects its plaintext. Each import tag is computed from this already-public ciphertext, the import's IV, and an independent destination key. Verification likewise depends only on the delivered ciphertext and tag and the destination channel state. GPU correction exposes no additional information: both its intermediate values and the recovered plaintext remain in protected GPU memory.

\bheading{Relation to ordinary CC transfers.} For the same source-key usage and permitted observations, rebinding inherits the confidentiality bound of ordinary D2H CC transfers, as quantified in \secref{sec:security-bound}. Repeated imports introduce no destination-key term into this bound.

\subsection{Integrity}
\label{sec:security-integrity}

\bheading{Property.} An import requested by trusted software either fails or recovers the plaintext of the export it names, except with a bounded probability of an authentication forgery.

\bheading{Argument.} Modifying the shared buffer cannot make the CPU authorize the modified ciphertext, because the CPU generates each import tag from the summary retained at export. This summary was computed from the same protected plaintext and D2H state as the CE export, independently of the shared buffer's contents.

The CE checks the resulting tag using the same GCM authentication equation as an ordinary H2D transfer (Equation~\eqref{eq:gcm-tag}). The check depends on the delivered ciphertext, the destination key, and the current IV, regardless of which key originally encrypted the data. Each fresh IV supplies a new pseudorandom tag mask, concealing the GHASH values and keeping observed tags from exposing the secret hash key. To make a different ciphertext pass, the adversary must choose a tag adjustment that matches the change in GHASH. For any fixed ciphertext modification and tag adjustment, this match holds for only a small fraction of possible hash keys. This is the forgery condition of ordinary GCM authentication, whose bound grows with message length and the number of verification attempts~\cite{gcm}.

Unless such a forgery succeeds, the CE has authenticated the intended ciphertext. Correction uses the source key and IV bound to that export to recover its original plaintext (Equation~\eqref{eq:retag-correction}). It operates on the CE output in protected GPU memory, and the application receives the result only after authentication and correction succeed.

\bheading{Relation to ordinary CC transfers.} Rebinding therefore preserves the ordinary H2D CC transfer's guarantee that successfully imported data matches the content authorized by trusted software. Giving existing ciphertext a new tag is an additional operation beyond standard GCM encryption. With re-tagging included in the destination key's usage, the authentication bound in Equation~\eqref{eq:security-bound} has the same form as for AES-GCM.

\subsection{Freshness}
\label{sec:security-freshness}

\bheading{Property.} Each import accepts only the ciphertext--tag pair authorized for its current channel IV, except with the authentication-forgery probability given in \secref{sec:security-bound}. Repeated imports of an existing export remain valid when trusted software issues a fresh tag for each transfer.

\bheading{Argument.} The CE verifies each import against its current trusted channel state, so an IV replayed from shared memory cannot select an earlier verification context. Every import obtains a fresh H2D IV through the channel's normal allocation path, shared with ordinary transfers under the same key. An allocated IV is never reused, even after a failed transfer.

For repeated imports of the same ciphertext under the same key, the GHASH summary is unchanged, but the tag mask changes with the IV. With 96-bit IVs and full-length tags, distinct IVs produce distinct masks because AES is a permutation, so the previous tag is rejected. Accepting any ciphertext--tag pair other than the one issued for the current import is the forgery analyzed in \secref{sec:security-integrity}.

\bheading{Relation to ordinary CC transfers.} As in ordinary CC transfers, replay protection relies on both GCM authentication and trusted IV progression~\cite{gcm}. Rebinding preserves this rule: the ciphertext can be retained, while each authorized import uses a fresh IV and a newly generated tag.

\subsection{Security Bound}
\label{sec:security-bound}

We analyze ordinary GCM operations together with fresh-IV re-tagging, with verification restricted to IVs already issued for transfers. For a key $K$, with 96-bit IVs and full 128-bit tags, the bound is
\begin{equation}
    \varepsilon_K = \varepsilon_{\mathrm{AES},K}
        + \frac{Q(Q-1)}{2^{129}}
        + \frac{q_v(\ell+1)}{2^{128}}.
    \label{eq:security-bound}
\end{equation}
The three terms capture AES security, accumulated key usage, and GHASH forgery attempts, as in standard AES-GCM bounds~\cite{gcm-proof-iom}. Here $\varepsilon_{\mathrm{AES},K}$ bounds the ability to distinguish AES from a random permutation, and $Q$ bounds the distinct AES inputs needed to generate observable encryption and tag outputs under $K$. In the forgery term, $q_v$ bounds verification attempts, and $\ell$ bounds the number of 128-bit ciphertext blocks in a message, excluding the length block. Ordinary encryption adds a payload-block input per 128-bit block and one tag-mask input to $Q$; each rebound import adds only one tag-mask input, since its intermediate plaintext remains protected. Ordinary traffic and re-tagging under the same key share this budget.

For a source and destination key, write these bounds as $\varepsilon_{\mathrm{src}}$ and $\varepsilon_{\mathrm{dst}}$. With permitted observations held fixed, an attacker identifying which of two equal-length candidate plaintexts was exported succeeds with probability at most $1/2+\varepsilon_{\mathrm{src}}$. The probability that any import accepts a ciphertext--tag pair other than the authorized one is at most $\varepsilon_{\mathrm{dst}}$. Correction correctness and trusted IV progression translate this authentication bound into the integrity and freshness guarantees above.

\bheading{Mechanized verification.} 
We machine-checked the correction identity and the confidentiality and integrity reductions in an abstract model of ciphertext rebinding with EasyCrypt~\cite{easycrypt}. The model considers one source and one destination key and includes ordinary traffic, repeated imports, and adaptive verification attempts. It uses abstract cryptographic primitives and assumes trusted channel state and metadata, secure internal transfers, and key-independent application data. The CUDA kernels, driver, and hardware are outside the verification.

\section{Case Studies}
\label{sec:case-studies}

We evaluate the performance of the ciphertext rebinding primitive through three case studies: model switching, KV-cache offloading, and cross-instance KV sharing. In each case study, the serving runtime retains and restores application state through the common export/import interface. 

\subsection{Model Switching}
\label{sec:case-model-switching}

Multi-model serving systems adjust GPU residency as demand changes to reduce the resources reserved for infrequently used models~\cite{aegaeon,serverlessllm}. Serving runtimes support releasing GPU weight memory while retaining a host copy for later restoration~\cite{vllmsleep}. Under CC, each restoration through the conventional transfer path still requires CPU encryption of the complete weight payload, even when the weights have not changed.

\sysname retains an exported ciphertext copy of the weights for repeated restoration. On the first release, it exports the loaded weights to shared host memory and keeps the associated metadata in CVM-private memory. While the weights and channel keys remain unchanged, subsequent releases require no further export. Each restoration imports the retained ciphertext with a fresh H2D IV, avoiding bulk CPU re-encryption. The export cost is amortized across restorations. A changed weight version needs a new export.

We integrate the same export/import interface into the weight-memory release and resume paths of SGLang~\cite{sglang} and vLLM~\cite{vllmsleep}. Their existing memory-management paths retain the weights' virtual addresses across release and resume, preserving references held by tensors and execution graphs. During release, the runtime waits for any required export to complete before freeing physical GPU memory. On resume, it restores the memory mappings and imports the retained ciphertext, returning only after the weights are ready for inference.

\subsection{KV Cache Offloading}
\label{sec:case-kv}

Multi-turn conversations and requests sharing long prefixes can reuse previously computed KV caches, but GPU memory cannot retain every useful prefix as the working set grows. Hierarchical caching systems such as CachedAttention and HiCache retain KV state beyond GPU memory, allowing later requests to restore cached prefixes and avoid repeated prefill~\cite{cachedattention,hicache}. Under CC, the conventional path decrypts each backup on the CPU and re-encrypts it for every restoration.

We integrate \sysname into SGLang HiCache's host-memory tier by mapping each cached page to exported ciphertext segments in shared host memory. The associated segment references and cryptographic metadata remain in CVM-private memory.

When HiCache backs up selected pages, their K and V regions across all layers are submitted for export. The GPU source pages remain valid and unchanged until export completes. Only then does the host copy become available for restoration, and the GPU source pages can be reclaimed. When a requested prefix is absent from GPU memory but present in the host cache, HiCache allocates destination GPU pages and imports the corresponding ciphertext. GPU correction and scatter place the recovered KV data into the destination layout. Attention can consume the restored pages only after import completes.

A retained host entry remains reusable while its contents and channel keys remain unchanged. Evicting a host entry removes its private references, and its storage can be reused only after outstanding transfers complete.

\subsection{Shared KV Pool across Instances}
\label{sec:case-shared-kv-pool}

To serve growing request demand, systems run multiple instances of the same model across GPUs. A shared KV cache in host memory or an external storage tier lets requests reuse prefixes computed by another instance, avoiding repeated prefill~\cite{dynamohicache}. Request routers such as Dynamo consider cache availability together with instance load when choosing where to execute a request~\cite{dynamorouting}. Under CC, the conventional path re-encrypts shared KV on the CPU for each restoration on a receiving instance.

We extend the HiCache integration to share exported KV state among SGLang instances serving the same model on different GPUs within one CVM. An exported copy serves cache hits on peer instances until its owner withdraws it to reclaim host capacity.

Instance $A$ populates the pool by exporting selected KV pages through the procedure in \secref{sec:retag-across-gpus}. It publishes the pages under their prefix-cache keys only after export completes. Dynamo uses publication and removal events from the pool to account for these shared prefixes when routing requests among the instances~\cite{dynamohicache}.

A shared-cache hit at instance $B$ creates a host-cache entry that references $A$'s exported pages, without copying the payload into another host buffer. Before importing the data, $B$ checks that the shared copy is still available and registers the transfer as active. It then imports the retained ciphertext into its allocated GPU pages using its own H2D channel. The transfer remains active until import is confirmed complete, after which attention can consume the restored KV data.

Under host-capacity pressure, $A$ withdraws a page from new imports. It sends invalidation notices to peer instances and reports the removal to the router. Withdrawal is serialized with the availability check and transfer registration: imports that register first may finish, while later attempts are treated as cache misses. Once active imports are confirmed complete, $A$ can reclaim the host storage even if $B$ still holds a host-cache reference to the withdrawn page. KV already restored into $B$'s GPU remains usable.

\section{Evaluation}
\label{sec:evaluation}

Our evaluation shows that by avoiding bulk CPU decryption and re-encryption, \sysname uses a small fraction of Multi-context CC's CPU time per GiB while approaching CC-off transfer bandwidth on H200. The transfer microbenchmarks also isolate the effects of batching and pipelining. These transfer gains bring model-loading latency from retained host copies close to CC-off in both SGLang and vLLM. We then examine their impact on serving throughput and latency with single-instance KV-cache offloading and cross-instance KV sharing. We also quantify memory overhead. Additional transfer and KV-cache offloading experiments on RTX PRO 6000 assess how the benefits vary across GPU platforms.

\subsection{Experimental Setup}
\label{sec:evaluation:setup}

\bheading{Implementation and testbed.} \sysname modifies NVIDIA's open GPU kernel modules~\cite{nvidiaopen} and provides a C++/CUDA library with serving adapters. Our main testbed is an AMD SEV-SNP~\cite{sevsnp} CVM with 16 vCPUs, 480~GiB of memory, and two H200 NVL GPUs on an AMD Zen~4 host. Each GPU connects to the host through PCIe~5.0 $\times$16. We also evaluate a 96~GB RTX PRO 6000 Blackwell Server Edition GPU on an AMD Zen~5 host, with 16 vCPUs assigned to the CVM. We use NVIDIA driver~610.57.04, CUDA~13.0, and PyTorch~2.13.0~\cite{pytorch}. Our serving integrations are based on SGLang~0.5.21~\cite{sglang} and vLLM~0.30.0, with Dynamo~1.6.1~\cite{dynamorouting} routing cross-instance requests.

\bheading{Compared systems.} \emph{Native CC} uses conventional CUDA transfers in one context under GPU CC, with CPU payload cryptography. \emph{Multi-context CC} follows Serialized Bridge~\cite{serializedbridge}, parallelizing these transfers across CPU workers and eight CUDA contexts per GPU. \sysname uses ciphertext export and import with batching and pipelining enabled by default. \emph{CC-off} provides a reference using native transfers with GPU CC disabled.

\bheading{KV-serving configuration.} Within each comparison, all paths use the same model, precision, and host-memory budget. We set SGLang's static GPU-memory fraction to 0.9 and hold each model's GPU-cache capacity fixed across paths. H200 KV experiments use a total host-cache budget of 460~GB, split equally between the two instances for shared KV. All paths use write-through backups and defer batch admission until required KV restores complete. Blocking restores run off the scheduler thread, following the same principle as asynchronous HiCache load submission~\cite{sglangasyncload}.

\bheading{KV-serving workloads.} For multi-turn reuse, we use a synthetic workload based on SGLang's HiCache multi-turn benchmark~\cite{sglangmultiturn}, denoted \emph{Multi-turn}. Each session has 10 turns, starting with 8,192 input tokens and adding 256 new input tokens per subsequent turn. Each turn generates 64 tokens, and later turns include the preceding inputs and generated responses. On H200, we use 240 sessions for Llama and Qwen and 100 for GLM. On RTX PRO 6000, we use 150 sessions for Llama. 

For shared-document reuse, following prior work on hierarchical KV caching~\cite{strata}, we use 105 documents from LooGLE's short-dependency QA subset~\cite{loogle} and 40 NarrativeQA test documents~\cite{narrativeqa}. Generation lengths follow the response texts provided by these datasets.

We sweep the offered request rate for all three workloads, with at most 128 requests in flight for single-instance serving and 256 for shared KV. LooGLE and NarrativeQA use Poisson arrivals. Multi-turn and LooGLE begin with empty caches. Following Strata~\cite{strata}, single-instance NarrativeQA starts with prepopulated host KV and an empty GPU cache.

\bheading{Measurement methodology.} Transfer bandwidth counts completed useful payload bytes per second. CPU cost sums user and kernel time across benchmark threads per GiB of useful payload. We report serving output-token throughput and mean time to first token (TTFT), measured from client dispatch to first-token receipt, over complete replays of all three workloads. Host-pool allocation and compilation warm-up precede measurement.

\subsection{Transfer Performance}
\label{sec:evaluation:transfer}

We measure sustained export (D2H) and import (H2D) bandwidth using 1~GiB contiguous buffers on H200 and RTX PRO 6000. Each configuration runs for a 2~s warm-up followed by a 10~s measurement window.

\begin{figure}[t]
\centering
\begingroup
\begin{tikzpicture}[font=\figurefont\fontsize{8}{9}\selectfont]
\begin{groupplot}[
group style={group size=2 by 2,horizontal sep=18pt,vertical sep=16pt},
scale only axis,width=0.39\columnwidth,height=.205\columnwidth,
axis line style={PaperInk!65},axis x line*=bottom,axis y line*=left,
tick align=outside,tick style={PaperInk!65},major tick length=1.6pt,
tick label style={font=\figurefont\fontsize{7}{8}\selectfont},
label style={font=\figurefont\fontsize{8}{9}\selectfont},
ylabel near ticks,ylabel style={inner sep=1pt},
xmin=.5,xmax=2.5,xtick={1,2},xticklabel style={align=center},
xticklabels={H200,{RTX PRO\\6000}},
scaled y ticks=false,clip=false,ymajorgrids=true,grid style={PaperInk!12},
legend columns=4,legend image post style={xscale=.6,yscale=.7},
legend style={draw=none,fill=none,font=\figurefont\fontsize{7}{8}\selectfont,
nodes={inner xsep=1pt},/tikz/every even column/.append style={column sep=2pt},row sep=0pt},
legend cell align=left
]
\nextgroupplot[ytick={0,20,40,60,80},xticklabels=\empty,ymin=0,ymax=80,ylabel={Bandwidth (GB/s)},legend to name=transfer-performance-legend]
\draw[PaperInk!45,line width=.55pt,dash pattern=on 2pt off 2.2pt] (rel axis cs:.5,0) -- (rel axis cs:.5,1);
\addlegendimage{area legend,draw=PaperProtected,fill=PaperProtected!75}
\addlegendentry{Native CC}
\addlegendimage{area legend,draw=PaperCipher,fill=PaperCipher!75}
\addlegendentry{Multi-context CC}
\addlegendimage{area legend,draw=PaperHighlight,fill=PaperHighlight!75}
\addlegendentry{\sysname}
\addlegendimage{area legend,draw=PaperInk!70,fill=PaperInk!35}
\addlegendentry{CC-off}
\addplot[forget plot,ybar,bar shift=0pt,bar width=8pt,draw=PaperProtected,fill=PaperProtected!75] coordinates {(0.640,9.66) (1.640,18.4)};
\addplot[forget plot,ybar,bar shift=0pt,bar width=8pt,draw=PaperCipher,fill=PaperCipher!75] coordinates {(0.880,29.31) (1.880,45.3)};
\addplot[forget plot,ybar,bar shift=0pt,bar width=8pt,draw=PaperHighlight,fill=PaperHighlight!75] coordinates {(1.120,54.43) (2.120,45)};
\addplot[forget plot,ybar,bar shift=0pt,bar width=8pt,draw=PaperInk!70,fill=PaperInk!35] coordinates {(1.360,56.37) (2.360,56)};
\node[anchor=south,font=\figurefont\fontsize{6}{7}\selectfont,text=PaperInk,fill=white,inner sep=.3pt,yshift=2.00pt] at (axis cs:0.640,9.66) {9.7};
\node[anchor=south,font=\figurefont\fontsize{6}{7}\selectfont,text=PaperInk,fill=white,inner sep=.3pt,yshift=2.00pt] at (axis cs:0.880,29.31) {29.3};
\node[anchor=south,font=\figurefont\fontsize{6}{7}\selectfont,text=PaperInk,fill=white,inner sep=.3pt,yshift=2.00pt] at (axis cs:1.120,54.43) {54.4};
\node[anchor=south,font=\figurefont\fontsize{6}{7}\selectfont,text=PaperInk,fill=white,inner sep=.3pt,yshift=2.00pt] at (axis cs:1.360,56.37) {56.4};
\node[anchor=south,font=\figurefont\fontsize{6}{7}\selectfont,text=PaperInk,fill=white,inner sep=.3pt,yshift=2.00pt] at (axis cs:1.640,18.4) {18.4};
\node[anchor=south,font=\figurefont\fontsize{6}{7}\selectfont,text=PaperInk,fill=white,inner sep=.3pt,yshift=2.00pt] at (axis cs:1.880,45.3) {45.3};
\node[anchor=south,font=\figurefont\fontsize{6}{7}\selectfont,text=PaperInk,fill=white,inner sep=.3pt,yshift=2.00pt] at (axis cs:2.120,45) {45.0};
\node[anchor=south,font=\figurefont\fontsize{6}{7}\selectfont,text=PaperInk,fill=white,inner sep=.3pt,yshift=2.00pt] at (axis cs:2.360,56) {56.0};
\nextgroupplot[ytick={0,20,40,60,80},xticklabels=\empty,ymin=0,ymax=80,yticklabels=\empty]
\draw[PaperInk!45,line width=.55pt,dash pattern=on 2pt off 2.2pt] (rel axis cs:.5,0) -- (rel axis cs:.5,1);
\addplot[forget plot,ybar,bar shift=0pt,bar width=8pt,draw=PaperProtected,fill=PaperProtected!75] coordinates {(0.640,8.48) (1.640,18.1)};
\addplot[forget plot,ybar,bar shift=0pt,bar width=8pt,draw=PaperCipher,fill=PaperCipher!75] coordinates {(0.880,24.48) (1.880,31.5)};
\addplot[forget plot,ybar,bar shift=0pt,bar width=8pt,draw=PaperHighlight,fill=PaperHighlight!75] coordinates {(1.120,56.58) (2.120,50.4)};
\addplot[forget plot,ybar,bar shift=0pt,bar width=8pt,draw=PaperInk!70,fill=PaperInk!35] coordinates {(1.360,56.48) (2.360,56.2)};
\node[anchor=south,font=\figurefont\fontsize{6}{7}\selectfont,text=PaperInk,fill=white,inner sep=.3pt,yshift=2.00pt] at (axis cs:0.640,8.48) {8.5};
\node[anchor=south,font=\figurefont\fontsize{6}{7}\selectfont,text=PaperInk,fill=white,inner sep=.3pt,yshift=2.00pt] at (axis cs:0.880,24.48) {24.5};
\node[anchor=south,font=\figurefont\fontsize{6}{7}\selectfont,text=PaperInk,fill=white,inner sep=.3pt,yshift=2.00pt] at (axis cs:1.120,56.58) {56.6};
\node[anchor=south,font=\figurefont\fontsize{6}{7}\selectfont,text=PaperInk,fill=white,inner sep=.3pt,yshift=2.00pt] at (axis cs:1.360,56.48) {56.5};
\node[anchor=south,font=\figurefont\fontsize{6}{7}\selectfont,text=PaperInk,fill=white,inner sep=.3pt,yshift=2.00pt] at (axis cs:1.640,18.1) {18.1};
\node[anchor=south,font=\figurefont\fontsize{6}{7}\selectfont,text=PaperInk,fill=white,inner sep=.3pt,yshift=2.00pt] at (axis cs:1.880,31.5) {31.5};
\node[anchor=south,font=\figurefont\fontsize{6}{7}\selectfont,text=PaperInk,fill=white,inner sep=.3pt,yshift=2.00pt] at (axis cs:2.120,50.4) {50.4};
\node[anchor=south,font=\figurefont\fontsize{6}{7}\selectfont,text=PaperInk,fill=white,inner sep=.3pt,yshift=2.00pt] at (axis cs:2.360,56.2) {56.2};
\nextgroupplot[ymode=log,log basis y=10,log origin y=infty,minor y tick num=0,ytick={1,10,100},yticklabels={1,10,100},ymin=0.5,ymax=800,ylabel={CPU (ms/GiB)}]
\draw[PaperInk!45,line width=.55pt,dash pattern=on 2pt off 2.2pt] (rel axis cs:.5,0) -- (rel axis cs:.5,1);
\addplot[forget plot,ybar,bar shift=0pt,bar width=8pt,draw=PaperProtected,fill=PaperProtected!75] coordinates {(0.640,109.4) (1.640,58.5)};
\addplot[forget plot,ybar,bar shift=0pt,bar width=8pt,draw=PaperCipher,fill=PaperCipher!75] coordinates {(0.880,257.1) (1.880,225.4)};
\addplot[forget plot,ybar,bar shift=0pt,bar width=8pt,draw=PaperHighlight,fill=PaperHighlight!75] coordinates {(1.120,8.1) (2.120,18)};
\addplot[forget plot,ybar,bar shift=0pt,bar width=8pt,draw=PaperInk!70,fill=PaperInk!35] coordinates {(1.360,1) (2.360,0.8)};
\node[anchor=south,font=\figurefont\fontsize{6}{7}\selectfont,text=PaperInk,fill=white,inner sep=.3pt,yshift=2.00pt] at (axis cs:0.640,109.4) {109};
\node[anchor=south,font=\figurefont\fontsize{6}{7}\selectfont,text=PaperInk,fill=white,inner sep=.3pt,yshift=2.00pt] at (axis cs:0.880,257.1) {257};
\node[anchor=south,font=\figurefont\fontsize{6}{7}\selectfont,text=PaperInk,fill=white,inner sep=.3pt,yshift=2.00pt] at (axis cs:1.120,8.1) {8.1};
\node[anchor=south,font=\figurefont\fontsize{6}{7}\selectfont,text=PaperInk,fill=white,inner sep=.3pt,yshift=2.00pt] at (axis cs:1.360,1) {1};
\node[anchor=south,font=\figurefont\fontsize{6}{7}\selectfont,text=PaperInk,fill=white,inner sep=.3pt,yshift=2.00pt] at (axis cs:1.640,58.5) {58.5};
\node[anchor=south,font=\figurefont\fontsize{6}{7}\selectfont,text=PaperInk,fill=white,inner sep=.3pt,yshift=2.00pt] at (axis cs:1.880,225.4) {225};
\node[anchor=south,font=\figurefont\fontsize{6}{7}\selectfont,text=PaperInk,fill=white,inner sep=.3pt,yshift=2.00pt] at (axis cs:2.120,18) {18};
\node[anchor=south,font=\figurefont\fontsize{6}{7}\selectfont,text=PaperInk,fill=white,inner sep=.3pt,yshift=2.00pt] at (axis cs:2.360,0.8) {0.8};
\nextgroupplot[ymode=log,log basis y=10,log origin y=infty,minor y tick num=0,ytick={1,10,100},yticklabels={1,10,100},ymin=0.5,ymax=800,yticklabels=\empty]
\draw[PaperInk!45,line width=.55pt,dash pattern=on 2pt off 2.2pt] (rel axis cs:.5,0) -- (rel axis cs:.5,1);
\addplot[forget plot,ybar,bar shift=0pt,bar width=8pt,draw=PaperProtected,fill=PaperProtected!75] coordinates {(0.640,127) (1.640,59.2)};
\addplot[forget plot,ybar,bar shift=0pt,bar width=8pt,draw=PaperCipher,fill=PaperCipher!75] coordinates {(0.880,284.5) (1.880,325.5)};
\addplot[forget plot,ybar,bar shift=0pt,bar width=8pt,draw=PaperHighlight,fill=PaperHighlight!75] coordinates {(1.120,7.8) (2.120,22.5)};
\addplot[forget plot,ybar,bar shift=0pt,bar width=8pt,draw=PaperInk!70,fill=PaperInk!35] coordinates {(1.360,1) (2.360,0.8)};
\node[anchor=south,font=\figurefont\fontsize{6}{7}\selectfont,text=PaperInk,fill=white,inner sep=.3pt,yshift=2.00pt] at (axis cs:0.640,127) {127};
\node[anchor=south,font=\figurefont\fontsize{6}{7}\selectfont,text=PaperInk,fill=white,inner sep=.3pt,yshift=2.00pt] at (axis cs:0.880,284.5) {285};
\node[anchor=south,font=\figurefont\fontsize{6}{7}\selectfont,text=PaperInk,fill=white,inner sep=.3pt,yshift=2.00pt] at (axis cs:1.120,7.8) {7.8};
\node[anchor=south,font=\figurefont\fontsize{6}{7}\selectfont,text=PaperInk,fill=white,inner sep=.3pt,yshift=2.00pt] at (axis cs:1.360,1) {1};
\node[anchor=south,font=\figurefont\fontsize{6}{7}\selectfont,text=PaperInk,fill=white,inner sep=.3pt,yshift=2.00pt] at (axis cs:1.640,59.2) {59.2};
\node[anchor=south,font=\figurefont\fontsize{6}{7}\selectfont,text=PaperInk,fill=white,inner sep=.3pt,yshift=2.00pt] at (axis cs:1.880,325.5) {326};
\node[anchor=south,font=\figurefont\fontsize{6}{7}\selectfont,text=PaperInk,fill=white,inner sep=.3pt,yshift=2.00pt] at (axis cs:2.120,22.5) {22.5};
\node[anchor=south,font=\figurefont\fontsize{6}{7}\selectfont,text=PaperInk,fill=white,inner sep=.3pt,yshift=2.00pt] at (axis cs:2.360,0.8) {0.8};
\end{groupplot}
\node[anchor=south,inner sep=0pt] at ([yshift=4pt]group c1r1.north) {(a) Export (D2H)};
\node[anchor=south,inner sep=0pt] at ([yshift=4pt]group c2r1.north) {(b) Import (H2D)};
\path (group c1r1.north west) -- node[midway,above=13pt,inner sep=0pt] {\pgfplotslegendfromname{transfer-performance-legend}} (group c2r1.north east);
\end{tikzpicture}
\endgroup
\caption{Transfer performance for 1~GiB payloads.
Top: bandwidth.
Bottom: total CPU time per GiB (log scale).}
\label{fig:transfer-performance}
\end{figure}
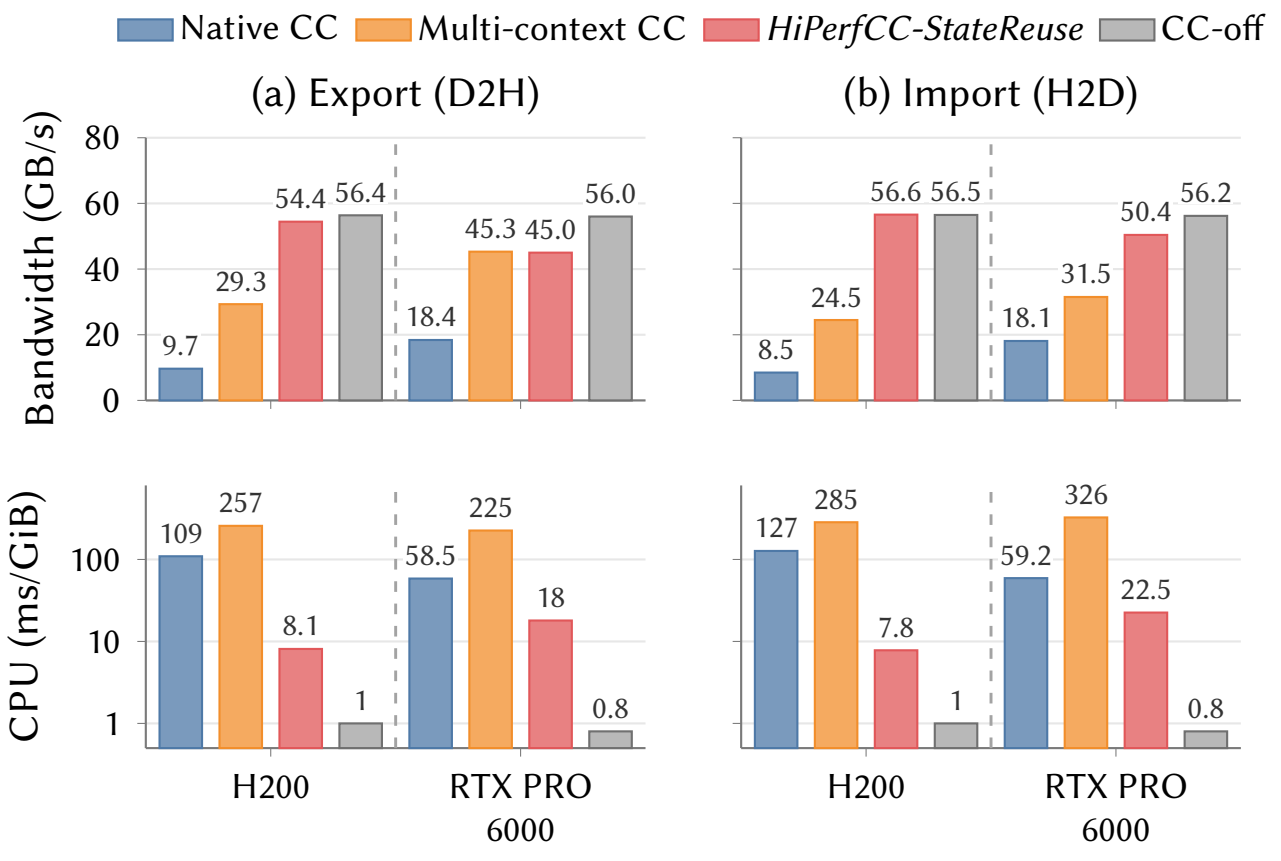

\bheading{Transfer bandwidth.} \figref{fig:transfer-performance} (top) shows that \sysname improves bandwidth over Native CC on both platforms and in both directions. On H200, it achieves 54--57~GB/s, approaching CC-off bandwidth and outperforming Multi-context CC by 1.9--2.3$\times$. On RTX PRO 6000, it matches Multi-context CC for export and improves import bandwidth by $1.6\times$, while retaining a larger gap to CC-off.

On RTX PRO 6000, faster CPU-side payload cryptography raises the bandwidth of Native CC and Multi-context CC, narrowing the relative gain from ciphertext rebinding. The Zen~5 host's full-width 512-bit vector execution likely contributes to the faster cryptography. The H200 host's Zen~4 cores instead execute 512-bit operations in two 256-bit steps~\cite{amdzenfour,amdzenfive}.

With CPU payload cryptography removed, CE parallelism becomes an important factor in \sysname's bandwidth. CC transfers require CEs that support hardware encryption and decryption~\cite{nvidiaopen}. On RTX PRO 6000, the driver exposes two such CEs, which we assign to export and import, respectively. H200 provides enough independent CEs to use two per direction. The lower CE parallelism on RTX PRO 6000 contributes to its remaining bandwidth gap to CC-off.

Gathering and scattering noncontiguous KV data costs little bandwidth: paged-buffer transfers stay within 3\% of contiguous ones on both platforms.

\bheading{CPU cost.} \sysname reduces CPU time per GiB by 92--97\% relative to Multi-context CC across both platforms and directions (\figref{fig:transfer-performance}, bottom). Multi-context CC distributes payload encryption and decryption across CPU workers, while ciphertext rebinding avoids that per-byte CPU work. Thus, even the comparable export bandwidth on RTX PRO 6000 comes with substantially lower CPU demand.

\label{sec:evaluation:ablation}

\bheading{Batching and pipelining.} With 1~GiB transfers on both platforms, the two optimizations in \secref{sec:pcie-speed} together increase bandwidth by 23--60\% over the configuration with neither enabled. \tabref{tab:ablation} reports bandwidth and CPU cost with staging allocation and CE counts held fixed.

\begin{table}[t]
\centering
\caption{Batching and pipelining ablation with 1~GiB transfers. BW: bandwidth (GB/s); CPU: CPU time (ms/GiB).}
\label{tab:ablation}
\begingroup
\fontsize{8}{10}\selectfont
\setlength{\tabcolsep}{2pt}
\renewcommand{\arraystretch}{1.15}
\begin{tabular*}{\columnwidth}{@{\extracolsep{\fill}}llrrrrrrrr@{}}
\toprule
 & & \multicolumn{2}{c}{Neither} & \multicolumn{2}{c}{Batching}
   & \multicolumn{2}{c}{Pipelining} & \multicolumn{2}{c}{Both} \\
\cmidrule(lr){3-4}\cmidrule(lr){5-6}\cmidrule(lr){7-8}\cmidrule(l){9-10}
GPU & Direction & BW & CPU & BW & CPU & BW & CPU & BW & CPU \\
\midrule
\multirow{2}{*}{H200}
 & Export & 34.0 & 19.1 & 42.8 & 5.7 & 34.9 & 18.9 & 54.4 & 8.1 \\
 & Import & 46.1 & 27.5 & 49.2 & 8.0 & 46.1 & 27.3 & 56.6 & 7.8 \\
\addlinespace[3pt]
\multirow{2}{*}{\shortstack[l]{RTX PRO\\6000}}
 & Export & 30.5 & 29.0 & 40.5 & 18.2 & 30.8 & 28.9 & 45.0 & 18.0 \\
 & Import & 37.7 & 23.2 & 38.8 & 22.3 & 47.2 & 23.2 & 50.4 & 22.5 \\
\bottomrule
\end{tabular*}
\endgroup
\end{table}

Batching raises bandwidth and lowers CPU cost by amortizing driver entry and protected command submission across segments. With pipelining disabled, batching increases export bandwidth by 26--33\% across the two platforms, with smaller gains for import. It also reduces CPU time per GiB by 70--71\% on H200.

Pipelining improves bandwidth mainly when batching is enabled, increasing it by 11--30\% across the two platforms and directions. Without batching, its only substantial gain is a 25\% improvement in RTX PRO 6000 import bandwidth. For export, individually submitted segments may finish transferring before the collected IVs reach the GPU for summary computation, leaving little work to overlap. Batching keeps more CE transfers outstanding when summary computation begins, enabling greater overlap. For import, H200 already overlaps GPU processing with transfers across its two CEs, though waits for staging reuse remain. RTX PRO 6000's single import CE must instead wait for correction and scatter to release its staging buffer when pipelining is disabled. Double buffering allows the next batch to arrive while GPU processing continues.

\subsection{Model Switching}
\label{sec:evaluation:model-switching}

\bheading{Setup.} We evaluate Llama-3.1-8B-Instruct (BF16) and, following Serialized Bridge~\cite{serializedbridge}, GPT-OSS-120B (MXFP4) on one H200 using SGLang and vLLM. After SGLang's full memory release or vLLM's level-1 sleep, we time the complete resume or wake-up call, including GPU memory allocation, mapping, and weight restoration.

\begin{figure}[t]
\centering
\begingroup
\begin{tikzpicture}[font=\figurefont\fontsize{8}{9}\selectfont]
\begin{groupplot}[
group style={group size=2 by 1,horizontal sep=12pt,vertical sep=16pt},
scale only axis,width=0.395\columnwidth,height=.25\columnwidth,
axis line style={PaperInk!65},axis x line*=bottom,axis y line*=left,
tick align=outside,tick style={PaperInk!65},major tick length=1.6pt,
tick label style={font=\figurefont\fontsize{7}{8}\selectfont},
label style={font=\figurefont\fontsize{8}{9}\selectfont},
ylabel near ticks,ylabel style={inner sep=1pt},
xmin=.5,xmax=2.5,xtick={1,2},xticklabel style={align=center},
xticklabels={SGLang,vLLM},
scaled y ticks=false,clip=false,ymajorgrids=true,grid style={PaperInk!12},
legend columns=4,legend image post style={xscale=.6,yscale=.7},
legend style={draw=none,fill=none,font=\figurefont\fontsize{7}{8}\selectfont,
nodes={inner xsep=1pt},/tikz/every even column/.append style={column sep=2pt},row sep=0pt},
legend cell align=left
]
\nextgroupplot[ytick={0,0.5,1,1.5,2,2.5},ymin=0,ymax=2.8,ylabel={Latency (s)},legend to name=model-switching-legend]
\draw[PaperInk!45,line width=.55pt,dash pattern=on 2pt off 2.2pt] (rel axis cs:.5,0) -- (rel axis cs:.5,1);
\addlegendimage{area legend,draw=PaperProtected,fill=PaperProtected!75}
\addlegendentry{Native CC}
\addlegendimage{area legend,draw=PaperCipher,fill=PaperCipher!75}
\addlegendentry{Multi-context CC}
\addlegendimage{area legend,draw=PaperHighlight,fill=PaperHighlight!75}
\addlegendentry{\sysname}
\addlegendimage{area legend,draw=PaperInk!70,fill=PaperInk!35}
\addlegendentry{CC-off}
\addplot[forget plot,ybar,bar shift=0pt,bar width=8pt,draw=PaperProtected,fill=PaperProtected!75] coordinates {(0.640,2.023) (1.640,2.233)};
\addplot[forget plot,ybar,bar shift=0pt,bar width=8pt,draw=PaperCipher,fill=PaperCipher!75] coordinates {(0.880,0.741) (1.880,1.074)};
\addplot[forget plot,ybar,bar shift=0pt,bar width=8pt,draw=PaperHighlight,fill=PaperHighlight!75] coordinates {(1.120,0.336) (2.120,0.665)};
\addplot[forget plot,ybar,bar shift=0pt,bar width=8pt,draw=PaperInk!70,fill=PaperInk!35] coordinates {(1.360,0.323) (2.360,0.726)};
\node[anchor=south,font=\figurefont\fontsize{6}{7}\selectfont,text=PaperInk,fill=white,inner sep=.3pt,yshift=2.00pt] at (axis cs:0.640,2.023) {2.02};
\node[anchor=south,font=\figurefont\fontsize{6}{7}\selectfont,text=PaperInk,fill=white,inner sep=.3pt,yshift=2.00pt] at (axis cs:0.880,0.741) {0.74};
\node[anchor=south,font=\figurefont\fontsize{6}{7}\selectfont,text=PaperInk,fill=white,inner sep=.3pt,yshift=2.00pt] at (axis cs:1.120,0.336) {0.34};
\node[anchor=south,font=\figurefont\fontsize{6}{7}\selectfont,text=PaperInk,fill=white,inner sep=.3pt,yshift=2.00pt] at (axis cs:1.360,0.323) {0.32};
\node[anchor=south,font=\figurefont\fontsize{6}{7}\selectfont,text=PaperInk,fill=white,inner sep=.3pt,yshift=2.00pt] at (axis cs:1.640,2.233) {2.23};
\node[anchor=south,font=\figurefont\fontsize{6}{7}\selectfont,text=PaperInk,fill=white,inner sep=.3pt,yshift=2.00pt] at (axis cs:1.880,1.074) {1.07};
\node[anchor=south,font=\figurefont\fontsize{6}{7}\selectfont,text=PaperInk,fill=white,inner sep=.3pt,yshift=2.00pt] at (axis cs:2.120,0.665) {0.67};
\node[anchor=south,font=\figurefont\fontsize{6}{7}\selectfont,text=PaperInk,fill=white,inner sep=.3pt,yshift=2.00pt] at (axis cs:2.360,0.726) {0.73};
\nextgroupplot[ytick={0,5,10,15},ymin=0,ymax=17]
\draw[PaperInk!45,line width=.55pt,dash pattern=on 2pt off 2.2pt] (rel axis cs:.5,0) -- (rel axis cs:.5,1);
\addplot[forget plot,ybar,bar shift=0pt,bar width=8pt,draw=PaperProtected,fill=PaperProtected!75] coordinates {(0.640,13.832) (1.640,7.981)};
\addplot[forget plot,ybar,bar shift=0pt,bar width=8pt,draw=PaperCipher,fill=PaperCipher!75] coordinates {(0.880,5.119) (1.880,3.024)};
\addplot[forget plot,ybar,bar shift=0pt,bar width=8pt,draw=PaperHighlight,fill=PaperHighlight!75] coordinates {(1.120,2.112) (2.120,1.418)};
\addplot[forget plot,ybar,bar shift=0pt,bar width=8pt,draw=PaperInk!70,fill=PaperInk!35] coordinates {(1.360,2.077) (2.360,1.602)};
\node[anchor=south,font=\figurefont\fontsize{6}{7}\selectfont,text=PaperInk,fill=white,inner sep=.3pt,yshift=2.00pt] at (axis cs:0.640,13.832) {13.83};
\node[anchor=south,font=\figurefont\fontsize{6}{7}\selectfont,text=PaperInk,fill=white,inner sep=.3pt,yshift=2.00pt] at (axis cs:0.880,5.119) {5.12};
\node[anchor=south,font=\figurefont\fontsize{6}{7}\selectfont,text=PaperInk,fill=white,inner sep=.3pt,yshift=2.00pt] at (axis cs:1.120,2.112) {2.11};
\node[anchor=south,font=\figurefont\fontsize{6}{7}\selectfont,text=PaperInk,fill=white,inner sep=.3pt,yshift=2.00pt] at (axis cs:1.360,2.077) {2.08};
\node[anchor=south,font=\figurefont\fontsize{6}{7}\selectfont,text=PaperInk,fill=white,inner sep=.3pt,yshift=2.00pt] at (axis cs:1.640,7.981) {7.98};
\node[anchor=south,font=\figurefont\fontsize{6}{7}\selectfont,text=PaperInk,fill=white,inner sep=.3pt,yshift=2.00pt] at (axis cs:1.880,3.024) {3.02};
\node[anchor=south,font=\figurefont\fontsize{6}{7}\selectfont,text=PaperInk,fill=white,inner sep=.3pt,yshift=2.00pt] at (axis cs:2.120,1.418) {1.42};
\node[anchor=south,font=\figurefont\fontsize{6}{7}\selectfont,text=PaperInk,fill=white,inner sep=.3pt,yshift=2.00pt] at (axis cs:2.360,1.602) {1.60};
\end{groupplot}
\node[anchor=south,inner sep=0pt] at ([yshift=4pt]group c1r1.north) {(a) Llama-3.1-8B};
\node[anchor=south,inner sep=0pt] at ([yshift=4pt]group c2r1.north) {(b) GPT-OSS-120B};
\path (group c1r1.north west) -- node[midway,above=13pt,inner sep=0pt] {\pgfplotslegendfromname{model-switching-legend}} (group c2r1.north east);
\end{tikzpicture}
\endgroup
\caption{Complete model restoration latency on H200.}
\label{fig:model-switching}
\end{figure}
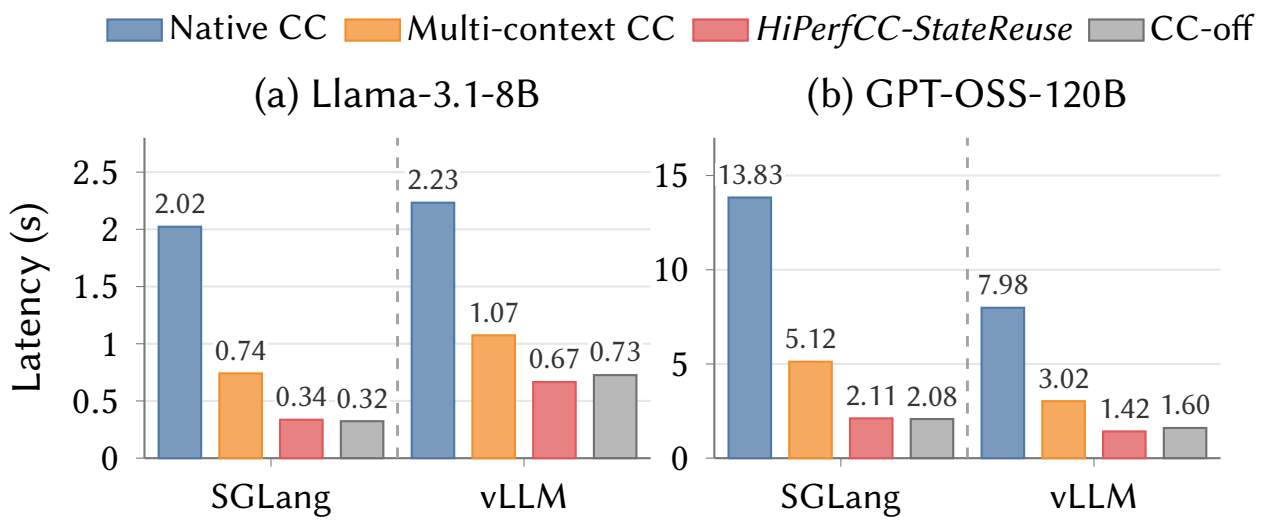

\bheading{Restoration performance.} \figref{fig:model-switching} shows that \sysname reduces complete restoration latency by 38--59\% relative to Multi-context CC across the two models and engines, bringing it close to the CC-off reference. For example, restoring GPT-OSS in SGLang takes 2.11~s, down from 5.12~s with Multi-context CC.

In vLLM, GPU allocation and mapping limit the reduction for the smaller Llama model, whose weights take less time to transfer. Recreating its allocations and mappings, including the KV-cache region, takes approximately 0.37~s, compared with 0.29~s for weight import. For the larger GPT-OSS model, weight transfer dominates restoration time, allowing faster imports to produce a larger reduction in complete wake-up latency.

\subsection{KV Cache Offloading}
\label{sec:evaluation:kv-offloading}

\input{figures/kv-offloading}

\bheading{Setup.} We evaluate SGLang HiCache on one H200 using Llama-3.1-8B-Instruct~\cite{llama31}, Qwen3.8-27B~\cite{qwen38}, and GLM-4.5-Air~\cite{glm45air} with a 4-bit AWQ checkpoint~\cite{glm45air-awq}. These models cover a dense Transformer with full attention (Llama), a hybrid of linear and full attention (Qwen), and a mixture-of-experts (MoE) Transformer (GLM). We use the Multi-turn, LooGLE, and NarrativeQA workloads described in \secref{sec:evaluation:setup}.

\bheading{Serving performance.} At the highest load evaluated for all four paths within each model--workload pair, \sysname achieves 1.38$\times$ Multi-context CC's throughput in geometric mean across the nine model--workload pairs and reduces mean TTFT by 28.8\% on average (\figref{fig:kv-offloading}). Faster KV restoration shortens the time requests wait before batch admission, improving TTFT and enabling higher throughput under heavier load.

Across workloads, NarrativeQA yields the largest throughput and TTFT gains for all three models. Its prepopulated host KV and short answers leave less new computation in the measured replay. Multi-turn and LooGLE include initial prefills from empty caches, while Multi-turn also processes new inputs at each turn and generates longer responses. This additional computation limits the end-to-end benefit of faster restoration and provides opportunities to overlap restores with other requests' GPU work.

GLM on Multi-turn illustrates this limited headroom: Native CC and \sysname both approach CC-off throughput, suggesting that restoration is not the dominant bottleneck. In this workload, later turns become eligible only after preceding turns complete, so execution timing also affects request interleaving and cache reuse. At the highest evaluated load, Multi-context CC restores about 27\% less KV from host memory than CC-off, helping explain why it outperforms both CC-off and \sysname.

Across models, Qwen shows the largest throughput gains over Multi-context CC on all three workloads. On NarrativeQA, for example, Multi-context CC provides less improvement over Native CC on Qwen than on Llama or GLM. SGLang's hybrid cache organizes host data by page while device KV is organized by layer. For Native CC and Multi-context CC, this layout mismatch splits KV restores into many small copies across pages and layers. The layout used for Llama and GLM allows adjacent pages to be combined into larger copies. Multiple contexts parallelize the small copies but retain their submission and synchronization overhead. \sysname batches ciphertext imports and restores the device layout on the GPU, reducing this overhead as well as CPU payload cryptography.

Several TTFT curves level off at high offered loads. Once the 128-request in-flight cap is reached, each new dispatch waits for an earlier request to complete. This bounds server-side concurrency and shifts additional waiting to the client, before TTFT measurement begins.

\subsection{Shared KV Pool across Instances}
\label{sec:evaluation:shared-kv-pool}

\bheading{Setup.} We evaluate two SGLang instances serving Llama-3.1-8B-Instruct in BF16, each on one H200 within the same CVM. All four paths enable cross-instance KV sharing and replay NarrativeQA~\cite{narrativeqa} through Dynamo~\cite{dynamorouting}. Before replay, each document is prefilled on one of the two instances and backed up to host memory; the resulting GPU-cache state is retained. The working set exceeds aggregate GPU-cache capacity but fits in the shared host pool.

\begin{figure}[t]
\centering
\begingroup
\pgfplotsset{
    shared native/.style={PaperProtected,solid,mark=*,mark size=1.7pt,line width=.8pt},
    shared contexts/.style={PaperCipher,solid,mark=square*,mark size=1.6pt,line width=.8pt},
    shared retag/.style={PaperHighlight,solid,mark=triangle*,mark size=2.1pt,line width=.9pt},
    shared off/.style={PaperInk!65,solid,mark=diamond*,mark size=1.7pt,line width=.8pt}
}
\begin{tikzpicture}[font=\figurefont\fontsize{8}{9}\selectfont]
\begin{groupplot}[
    group style={group name=shared,group size=2 by 1,horizontal sep=28pt},
    scale only axis,width=.36\columnwidth,height=.205\columnwidth,
    axis line style={PaperInk!65},tick style={PaperInk!65},
    tick align=outside,tick pos=left,major tick length=1.6pt,minor tick length=0pt,
    tick label style={font=\figurefont\fontsize{7}{8}\selectfont},
    label style={font=\figurefont\fontsize{8}{9}\selectfont},
    scaled ticks=false,/pgf/number format/fixed,/pgf/number format/1000 sep={},
    ylabel near ticks,ylabel style={inner sep=1pt},
    xmin=0,xmax=170,xtick={0,32,64,96,128,160},ymin=0,
    ymajorgrids=true,grid style={PaperInk!12},
    legend columns=4,legend cell align=left,
    legend image post style={xscale=.6,yscale=.7},
    legend style={draw=none,fill=none,font=\figurefont\fontsize{7}{8}\selectfont,
        nodes={inner xsep=1pt},
        /tikz/every even column/.append style={column sep=2pt},row sep=0pt}
]
\nextgroupplot[ymax=700,ytick={0,200,400,600},ylabel={Output tok/s},legend to name=shared-kv-pool-legend]
\addlegendimage{shared native}
\addlegendentry{Native CC}
\addlegendimage{shared contexts}
\addlegendentry{Multi-context CC}
\addlegendimage{shared retag}
\addlegendentry{\sysname}
\addlegendimage{shared off}
\addlegendentry{CC-off}
\addplot[shared off,forget plot] coordinates {(32,178.8) (64,340.7) (96,472.5) (128,609.3) (160,613)};
\addplot[shared native,forget plot] coordinates {(32,124.8) (64,162) (96,196.3) (128,189.2) (160,211.3)};
\addplot[shared contexts,forget plot] coordinates {(32,161.9) (64,274.3) (96,309.7) (128,389.9) (160,408.6)};
\addplot[shared retag,forget plot] coordinates {(32,177.3) (64,319.1) (96,453.4) (128,548.6) (160,564.9)};
\node[anchor=north west,inner sep=0pt] at (rel axis cs:.035,.95) {(a)};
\nextgroupplot[ymax=7,ytick={0,2,4,6},ylabel={Mean TTFT (s)}]
\addplot[shared off,forget plot] coordinates {(32,0.3565672466) (64,0.6743442406) (96,0.8321798845) (128,0.7190611486) (160,1.134833517)};
\addplot[shared native,forget plot] coordinates {(32,5.625748238) (64,5.381374727) (96,4.103658566) (128,4.84748877) (160,4.085449262)};
\addplot[shared contexts,forget plot] coordinates {(32,1.963758149) (64,2.680432607) (96,2.759098073) (128,2.120983775) (160,1.774938659)};
\addplot[shared retag,forget plot] coordinates {(32,0.5814469215) (64,1.00721109) (96,0.999623266) (128,0.9375937509) (160,1.215904496)};
\node[anchor=north west,inner sep=0pt] at (rel axis cs:.035,.95) {(b)};
\end{groupplot}
\node[anchor=south east,align=right,inner sep=0pt,text=PaperInk,font=\figurefont\fontsize{7}{8}\selectfont] at ([yshift=3pt]shared c1r1.north east) {MC \textbf{+38.3\%}\enspace|\enspace CC-off \textbf{-7.8\%}};
\node[anchor=south east,align=right,inner sep=0pt,text=PaperInk,font=\figurefont\fontsize{7}{8}\selectfont] at ([yshift=3pt]shared c2r1.north east) {MC \textbf{-31.5\%}\enspace|\enspace CC-off \textbf{+7.1\%}};
\path (shared c1r1.north west) -- node[midway,above=11pt,inner sep=0pt] {\pgfplotslegendfromname{shared-kv-pool-legend}} (shared c2r1.north east);
\path (shared c1r1.south west) -- node[midway,below=11pt] {Total offered request rate (req/s)} (shared c2r1.south east);
\end{tikzpicture}
\endgroup
\caption{Two Llama-3.1-8B instances on two H200 GPUs, with KV sharing enabled. Throughput totals both instances.}
\label{fig:shared-kv-pool}
\end{figure}
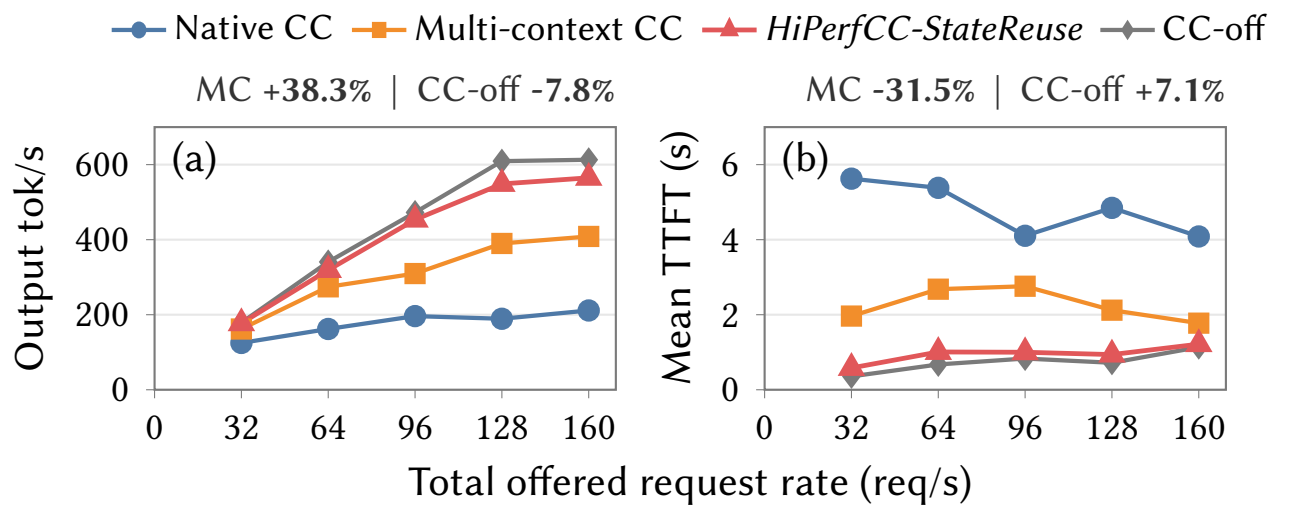

\bheading{Serving performance.} At a total offered request rate of 160 requests/s, \sysname achieves 1.38$\times$ Multi-context CC's throughput and reduces mean TTFT by 31.5\%, reaching 92.2\% of CC-off throughput (\figref{fig:shared-kv-pool}).

All paths support cross-instance KV reuse, so these gains demonstrate the benefit of accelerating restoration when requests consume KV produced by another GPU. \sysname reuses retained ciphertext across GPU channels, avoiding the CPU re-encryption that Multi-context CC still performs on each restoration.

\sysname maintains lower TTFT than Multi-context CC throughout the load sweep, although absolute TTFT varies non-monotonically with load. Our execution logs show that higher-load runs restore less KV from host memory for the same request trace. Concurrent requests sharing a document prefix on the same instance reuse the restored KV, allowing one restoration to serve multiple requests and reducing restoration work per request.

\subsection{Memory Overhead}
\label{sec:evaluation:overheads}

\sysname uses reusable GPU buffers and CVM-private host metadata. Our configuration allocates 1~GiB of GPU staging for each transfer direction, totaling 2~GiB per GPU.

For each 2~MiB ciphertext segment, trusted metadata retains its 12-byte D2H IV and a 16-byte authentication summary $g$ for each distinct destination H2D key, covering candidate import channels on both the same GPU and other GPUs. In our single-GPU H200 configuration, each library segment record occupies 112~bytes, corresponding to about 0.055~MiB per GiB of ciphertext capacity.

\subsection{Generality Across GPU Platforms}
\label{sec:evaluation:generality}

\bheading{Setup.} We repeat the KV-cache offloading experiments with Llama-3.1-8B-Instruct in BF16 on one RTX PRO 6000, using SGLang HiCache and the three workloads described in \secref{sec:evaluation:setup}.

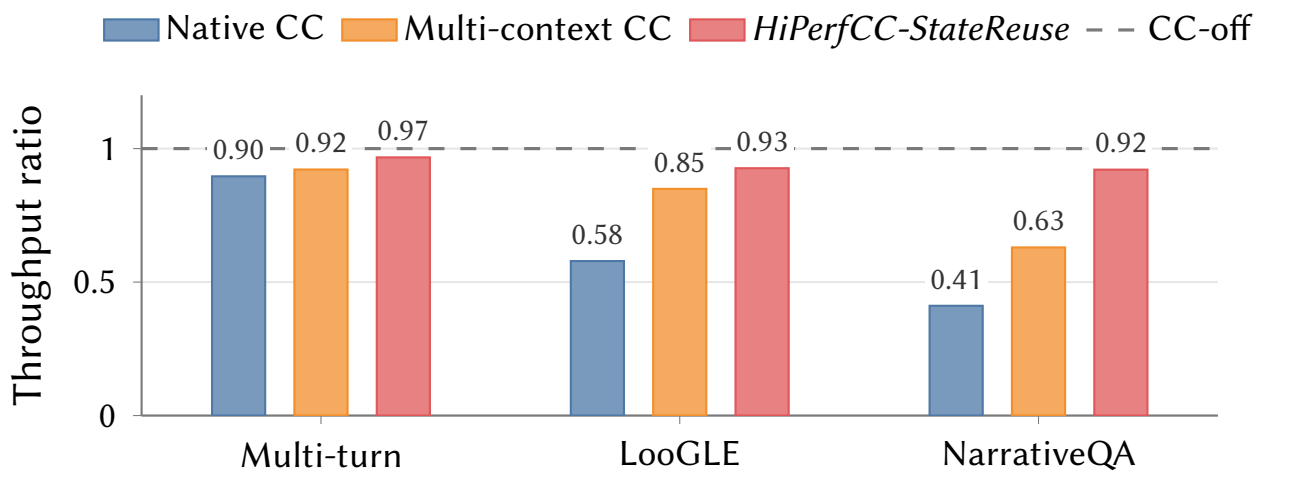
\begin{figure}[t]
\centering
\begingroup
\begin{tikzpicture}[font=\figurefont\fontsize{8}{9}\selectfont]
\begin{groupplot}[
    group style={group size=1 by 1},
    scale only axis,
    width=.84\columnwidth,height=.25\columnwidth,
    axis line style={PaperInk!65},axis x line*=bottom,axis y line*=left,
    tick align=outside,tick style={PaperInk!65},major tick length=1.6pt,
    tick label style={font=\figurefont\fontsize{7}{8}\selectfont},
    label style={font=\figurefont\fontsize{8}{9}\selectfont},
    ylabel near ticks,ylabel style={inner sep=1pt},
    xmin=-.5,xmax=2.5,xtick={0,1,2},
    xticklabels={Multi-turn,LooGLE,NarrativeQA},
    ymin=0,scaled ticks=false,
    ymajorgrids=true,grid style={PaperInk!12},
    legend columns=4,
    legend image post style={xscale=.6,yscale=.7},
    legend style={draw=none,fill=none,
        font=\figurefont\fontsize{7}{8}\selectfont,
        nodes={inner xsep=1pt},
        /tikz/every even column/.append style={column sep=2pt},row sep=0pt},
    legend cell align=left
]
\nextgroupplot[ylabel={Throughput ratio},ymax=1.2,ytick={0,0.5,1},yticklabels={0,0.5,1},legend to name=pro6000-kv-offloading-legend]
\addlegendimage{area legend,draw=PaperProtected,fill=PaperProtected!75}
\addlegendentry{Native CC}
\addlegendimage{area legend,draw=PaperCipher,fill=PaperCipher!75}
\addlegendentry{Multi-context CC}
\addlegendimage{area legend,draw=PaperHighlight,fill=PaperHighlight!75}
\addlegendentry{\sysname}
\addlegendimage{PaperInk!65,dashed,line width=.6pt}
\addlegendentry{CC-off}
\addplot[forget plot,ybar,bar width=10pt,bar shift=-15.5pt,draw=PaperProtected,fill=PaperProtected!75,point meta=explicit symbolic,nodes near coords={\pgfplotspointmeta},every node near coord/.append style={font=\figurefont\fontsize{6}{7}\selectfont,text=PaperInk,fill=white,inner sep=1pt,xshift=-15.5pt,yshift=2pt}]
    coordinates {(0,0.896203988)[0.90] (1,0.578753063)[0.58] (2,0.411210032)[0.41]};
\addplot[forget plot,ybar,bar width=10pt,bar shift=0pt,draw=PaperCipher,fill=PaperCipher!75,point meta=explicit symbolic,nodes near coords={\pgfplotspointmeta},every node near coord/.append style={font=\figurefont\fontsize{6}{7}\selectfont,text=PaperInk,fill=white,inner sep=1pt,xshift=0pt,yshift=2pt}]
    coordinates {(0,0.921664964)[0.92] (1,0.849094324)[0.85] (2,0.629798255)[0.63]};
\addplot[forget plot,ybar,bar width=10pt,bar shift=15.5pt,draw=PaperHighlight,fill=PaperHighlight!75,point meta=explicit symbolic,nodes near coords={\pgfplotspointmeta},every node near coord/.append style={font=\figurefont\fontsize{6}{7}\selectfont,text=PaperInk,fill=white,inner sep=1pt,xshift=15.5pt,yshift=2pt}]
    coordinates {(0,0.967101190)[0.97] (1,0.926594968)[0.93] (2,0.921204207)[0.92]};
\addplot[forget plot,PaperInk!65,dashed,line width=.6pt] coordinates {(-.5,1) (2.5,1)};
\end{groupplot}
\node[anchor=south,inner sep=0pt] at ([yshift=5pt]group c1r1.north) {\pgfplotslegendfromname{pro6000-kv-offloading-legend}};
\end{tikzpicture}
\endgroup
\caption{KV offloading with Llama-3.1-8B on RTX PRO 6000. Throughput is normalized to CC-off.}
\label{fig:pro6000-kv-offloading}
\end{figure}

\bheading{Serving performance.} At a common load near CC-off saturation in each workload, \sysname achieves 1.19$\times$ Multi-context CC's throughput and reaches 93.8\% of CC-off throughput, both in geometric mean across the three workloads. It reduces mean TTFT by 28.2\% on average relative to Multi-context CC. \figref{fig:pro6000-kv-offloading} reports throughput normalized to CC-off. NarrativeQA shows the largest throughput gains over Multi-context CC, consistent with the trend on H200.

As shown in \secref{sec:evaluation:transfer}, faster CPU-side cryptography on the RTX PRO 6000 host strengthens the conventional-transfer baselines, while the single CE limits \sysname's import bandwidth. These differences narrow its transfer advantage relative to H200.

\section{Discussion and Limitations}
\label{sec:discussion}

\bheading{Host-memory sharing and availability.} Denial-of-service (DoS) attacks are outside the conventional CPU--GPU CC threat model adopted in this work~\cite{gpucc,sevsnp}. Preparing the ciphertext pool requires converting host pages from CVM-private to shared memory for GPU DMA. Under SEV-SNP, the guest coordinates these changes with the hypervisor~\cite{sevsnp}. The shared pages hold GPU-exported ciphertext, while keys, summaries, and the metadata binding each segment to its export remain protected inside the private memory. As analyzed in \secref{sec:security-integrity}, the CPU derives tags from private summaries, so altering shared ciphertext cannot authorize the altered content, only causing DoS attack. Even with ordinary CC transfers, a malicious host can withhold resources or disrupt transfers to cause availability issues.  

\label{sec:discussion:kv-leakage}

\bheading{Access-pattern leakage and security scope.}
Side-channel attacks and access-pattern leakage are outside the conventional CPU--GPU CC threat model adopted in this work~\cite{gpucc,sevsnp}. Existing CC systems can expose application access patterns even when payload confidentiality and transfer authentication are preserved. For example, shared KV caches can reveal users' cached prompt prefixes through response timing and service order~\cite{earlybird,promptpeek}, while an untrusted host can observe application access traces through page-level and microarchitectural side channels on SEV-SNP and TDX~\cite{sevstep,tdxray}.

In our design, repeated imports of the same exported KV segment can also be recognized by the host because the retained ciphertext is stored in shared memory and remains unchanged despite fresh IVs and authentication tags. If the host has associated an exported segment with a known prompt prefix, recognizing subsequent imports may reveal accesses to that prefix. Although unchanged ciphertext in shared memory can simplify access-pattern leakage, it does not compromise payload encryption or reveal plaintext equality between independently encrypted segments. Preventing such access-pattern leakage requires additional mechanisms to conceal cache reuse and observable application behavior, a broader challenge for TEE-based systems that is beyond the scope of this work.

\label{sec:discussion:key-lifecycle}

\bheading{Key rotation and retained-state lifecycle.} Re-tagging consumes far less key budget per byte than ordinary transfers. Long-running services rotate channel keys because the security bound grows as more data is encrypted under the same key. Of the three terms in Equation~\eqref{eq:security-bound}, the AES term is assumed negligible, while full-length tags and termination on authentication failure keep the forgery term small. The remaining quadratic key-usage term $Q(Q-1)/2^{129}$ grows with cumulative encryption work and therefore determines the traffic-based usage limit. NVIDIA's driver sets its rotation threshold from this term~\cite{nvidiaopen}, with a default budget of $2^{-60}$.

With 2~MiB segments, this budget corresponds to approximately 390~GB of ordinary transfers per key. Rebinding uses one AES evaluation to generate a 16-byte tag mask for each 2~MiB import. The same budget therefore permits approximately 390~GB of tag masks, supporting about 51~PB of cumulative imports when the key is used only for re-tagging. Ordinary transfers and protected metadata under the same key consume part of this budget, reducing the rebinding volume supported before rotation.

Retaining ciphertext across rotations requires preserving the state needed to recover it and updating the summaries used to authenticate its imports. D2H rotation changes the key for new exports, while existing ciphertexts remain recoverable with their original keys and IVs. The library retains source-key versions needed by live segments or in-flight operations and reclaims obsolete versions. H2D rotation leaves the ciphertext unchanged but requires new authentication summaries $g$ under the destination key. When notified that an H2D key is nearing its limit, the library imports retained data through the authenticated path and computes the new summaries on the GPU in the background. Once the summaries are ready, it requests the key switch and uses the matching summaries for subsequent imports. If the driver rotates the key before this preparation completes, entries without a new summary become unavailable for import.

In separate measurements of D2H key rotation and H2D summary refresh, all retained segments remained importable. On H200, we estimate that D2H key rotation adds approximately 130~ms to the gaps between completed 1~GiB exports, after subtracting the steady-state completion interval. In an isolated measurement, refreshing $g$ for about 430~GiB of retained data took approximately 15~s per H2D key.

In restoration-heavy workloads, ciphertext rebinding handles much of the H2D payload traffic, slowing consumption of the H2D key budget and reducing how often summaries need to be refreshed. In the H200 KV-cache experiments, cumulative ciphertext imports reached approximately 9.3~TB per replay, less than 0.02\% of the 51~PB re-tagging capacity of a single key. We observed D2H key rotations, but no H2D key rotations or summary refreshes in these runs.

\section{Related Work}
\label{sec:related-work}

\bheading{Confidential GPU data movement.} Prior studies characterize transfer and runtime overheads in confidential GPU workloads~\cite{dissect_cc,ispass_cc}. Several designs reduce protected-transfer overhead but require hardware or GPU CC protocol changes unsupported by existing confidential GPUs. \laegis proposes GPU extensions with an HBM-resident IV bank, decoupling CPU pre-encryption from UVM page-migration order~\cite{laegis}. TensorTEE, evaluated in simulation, unifies the protection granularity of CPU and NPU TEEs at the tensor level, enabling data transfer between them without re-encryption~\cite{2024:tensortee:han}. Fastrack proposes changes to the GPU CC protocol that establish a direct client--GPU secure channel, avoiding CPU decryption and re-encryption of client inputs~\cite{fastrack}. AMD SEV-TIO and Intel TDX Connect provide trusted I/O, letting trusted devices directly access CVM-private memory without shared host buffers, given compatible hosts, devices, firmware, and software~\cite{amdsevtio,tdxconnect}.

In contrast, optimizations on existing confidential GPUs mainly overlap or parallelize CPU cryptography, retaining its per-byte cost for GPU state backup and restoration. PipeLLM predicts transfer buffers and IVs to overlap CPU encryption with GPU computation~\cite{pipellm}. Serialized Bridge uses persistent CUDA context pools to address serialization in confidential CPU--GPU model loading~\cite{serializedbridge}.

Retaining ciphertext offers another way to reduce repeated cryptographic work. PipeLLM discusses ciphertext reuse for unchanged swap data as an opportunity for future hardware and library support~\cite{pipellm}. FLASH retains GPU-produced ciphertext in checkpoints, avoiding CPU decryption and re-encryption during checkpoint creation~\cite{flash}. Its recovery path decrypts the saved GPU state inside the CVM and restores it through conventional protected H2D transfers. Ciphertext rebinding instead imports retained ciphertext through existing CE interfaces, adapting authentication and plaintext recovery to the destination channel's key and fresh IVs. \sysname uses this primitive for repeated restoration and cross-GPU sharing within a CVM trust domain, avoiding bulk CPU decryption and re-encryption without GPU hardware or firmware changes.

\bheading{Model and KV-state management.} Aegaeon pools GPUs for concurrent model serving, while ServerlessLLM optimizes model loading, placement, and migration~\cite{aegaeon,serverlessllm}. CachedAttention and HiCache organize KV reuse across memory and storage tiers~\cite{cachedattention,hicache}. Strata addresses fragmented KV transfers with GPU-assisted I/O and cache-aware scheduling~\cite{strata}. Mooncake and Dynamo coordinate distributed KV reuse with request scheduling~\cite{mooncake,dynamorouting}. \sysname complements these placement, caching, and scheduling techniques by reducing the CPU cryptographic cost of transferring retained GPU state under CC.

\section{Conclusion}
\label{sec:conclusion}

\sysname implements ciphertext rebinding for repeated and cross-GPU ciphertext reuse within a CVM trust domain. Retained authentication summaries enable CPU tag updates without payload processing, while GPU correction recovers the original plaintext after CE decryption under fresh channel state. We analyzed its confidentiality, integrity, and channel freshness under the stated threat model. Our batched, pipelined driver and library require no GPU hardware or firmware changes. Compared with Multi-context CC, \sysname reduces CPU time per GiB by 92--97\% across H200 and RTX PRO 6000 and achieves 1.9--2.3$\times$ transfer bandwidth on H200. On H200, it reduces model-restoration latency by 38--59\%, achieves 1.38$\times$ the baseline's geometric-mean output-token throughput with KV offloading, and reaches 1.38$\times$ the baseline's throughput with cross-instance KV sharing.

\bibliographystyle{ACM-Reference-Format}
\bibliography{ref}

\end{document}